\documentclass[aps,prd,preprintnumbers,groupedaddress,nofootinbib,amssymb,notitlepage,eqsecnum]{revtex4-1}
\usepackage{here}
\usepackage[dvipdfmx]{graphicx}
\usepackage{amsmath,amsthm,amssymb}
\usepackage{bm}
\usepackage{color}

\usepackage{amsfonts}
\usepackage{dcolumn}
\usepackage{hyperref}
\allowdisplaybreaks[1]
\usepackage{stackengine}

\newcommand{\be}{\begin{equation}}  
\newcommand{\ee}{\end{equation}}
\newcommand{\ba}{\begin{eqnarray}}
\newcommand{\ea}{\end{eqnarray}}

\newcommand{\rd}{{\rm d}}

\newcommand{\bem}{\begin{bmatrix}}
\newcommand{\eem}{\end{bmatrix}}
\newcommand{\Mpl}{M_{\rm Pl}}

\allowdisplaybreaks

\begin{document}

\preprint{WUCG-22-05}
\title{Stability of neutron stars in Horndeski theories with Gauss-Bonnet couplings}

\author{Masato Minamitsuji$^{1}$ and Shinji Tsujikawa$^{2}$}

\affiliation{
$^{1}$Centro de Astrof\'{\i}sica e Gravita\c c\~ao - CENTRA, Departamento de F\'{\i}sica, Instituto Superior T\'ecnico - IST, Universidade de Lisboa - UL, Av.~Rovisco Pais 1, 1049-001 Lisboa, Portugal\\
$^2$Department of Physics, Waseda University, 3-4-1 Okubo, Shinjuku, Tokyo 169-8555, Japan}

\begin{abstract}

In Horndeski theories containing a scalar coupling with the Gauss-Bonnet (GB) curvature invariant $R_{\rm GB}^2$, we study the existence and linear stability of neutron star (NS) solutions on a static and spherically symmetric background. For a scalar-GB coupling of the form $\alpha \xi(\phi) R_{\rm GB}^2$, where $\xi$ is a function of the scalar field $\phi$, the existence of linearly stable stars with a nontrivial scalar profile without instabilities puts an upper bound on the strength of the dimensionless coupling constant $|\alpha|$. To realize maximum masses of NSs for a linear (or dilatonic) GB coupling $\alpha_{\rm GB}\phi R_{\rm GB}^2$ with typical nuclear equations of state, we obtain the theoretical upper limit $\sqrt{|\alpha_{\rm GB}|}<0.7~{\rm km}$. This is tighter than those obtained by the observations of gravitational waves emitted from binaries containing NSs. We also incorporate cubic-order scalar derivative interactions, quartic derivative couplings with nonminimal couplings to a Ricci scalar besides the scalar-GB coupling and show that NS solutions with a nontrivial scalar profile satisfying all the linear stability conditions are present for certain ranges of the coupling constants. In regularized 4-dimensional Einstein-GB gravity obtained from a Kaluza-Klein reduction with an appropriate rescaling of the GB coupling constant, we find that NSs in this theory suffer from a strong coupling problem as well as Laplacian instability of even-parity perturbations. We also study NS solutions with a nontrivial scalar profile in power-law $F(R_{\rm GB}^2)$ models, and show that they are pathological in the interior of stars and plagued by ghost instability together with the asymptotic strong coupling problem in the exterior of stars.
\end{abstract}

\date{\today}


\maketitle

\section{Introduction}
\label{introsec}

After the dawn of gravitational waves (GW) astronomy 
from a binary system of black holes (BHs) \cite{LIGOScientific:2016aoc}, a new observational probe 
of the physics on a strong gravitational background
has begun. 
In particular, the GW170817 event \cite{LIGOScientific:2017vwq}
allowed us to put constraints on the mass-radius relation of 
neutron stars (NSs) from their tidal deformation before 
the coalescence \cite{LIGOScientific:2018cki}. 
After a merger of binaries, a compact object exhibits 
a damped sinusoidal oscillation with quasi-normal 
frequencies \cite{Kokkotas:1999bd,Nollert:1999ji}.  
Upcoming observational data of GWs will provide us further 
detailed information for new physics in strong gravity regimes
in the vicinity of BHs and in the interior of NSs.

{}From cosmological observational data, we know that 
about 95\,\% of the energy density of today's Universe 
is dominated by mysterious components dubbed dark energy and 
dark matter \cite{SupernovaSearchTeam:1998fmf,SupernovaCosmologyProject:1998vns,WMAP:2003elm,SDSS:2003eyi,Planck:2013pxb}. 
Since it is still challenging to explain their origins 
within the framework of General Relativity (GR) and 
Standard Model of particle physics, there is 
a motivation to introduce new degrees of freedom \cite{Copeland:2006wr,Silvestri:2009hh,DeFelice:2010aj,Clifton:2011jh,Joyce:2014kja,Koyama:2015vza,Heisenberg:2018vsk,Kase:2018aps}. 
A scalar field is one of the simplest candidates,
and widely used to explain physical phenomena relevant to 
the dark sector of the Universe.

On a spherically symmetric background, 
an asymptotically-flat vacuum solution in GR is
uniquely described by the Schwarzschild metric 
with the mass of a compact body.
The background Schwarzschild geometry can be modified 
by introducing a new degree of freedom. 
For asymptotically-flat BHs, however, the property of the 
absence of a nontrivial scalar profile holds for 
a wide class of scalar-tensor theories--including 
a canonical scalar 
field \cite{Hawking:1971vc,Bekenstein:1972ny}, 
k-essence \cite{Graham:2014mda}, 
nonminimally coupled scalar field with Ricci 
scalar \cite{Hawking:1972qk,Bekenstein:1995un,Sotiriou:2011dz,Faraoni:2017ock}, 
and regular derivative couplings 
in shift-symmetric Horndeski theories \cite{Hui:2012qt}. 
An exceptional case is a scalar coupling with the 
GB curvature invariant $R_{\rm GB}^2$ of the form 
$\alpha \xi(\phi) R_{\rm GB}^2$, where $\alpha$ is a dimensionless 
coupling constant and $\xi$ is a regular function of the scalar
field $\phi$ \cite{Kanti:1995vq,Torii:1996yi,Kanti:1997br,Chen:2006ge,Guo:2008hf,Guo:2008eq,Pani:2009wy,Sotiriou:2013qea,Sotiriou:2014pfa,Ayzenberg:2014aka,Maselli:2015tta,Kleihaus:2011tg,Kleihaus:2015aje,Doneva:2017bvd,Silva:2017uqg,Antoniou:2017acq,Blazquez-Salcedo:2018jnn,Minamitsuji:2018xde,Silva:2018qhn,Macedo:2019sem,Doneva:2021tvn,Minamitsuji:2022vbi}. 
This scalar-GB coupling belongs to a subclass of Horndeski 
theories \cite{Horndeski} containing nonanalytic functions like $\ln |X|$ 
in the coupling functions $G_{2,3,4,5}$ \cite{KYY}, where $X$ is 
a canonical field kinetic term.

For a NS, the presence of matter inside the star can modify 
the no-hair property of BHs in several subclasses of 
Horndeski theories. 
Nonminimal couplings with the Ricci scalar $R$ of the form 
$G_4(\phi)R$ allow a possibility for realizing NS solutions
endowed with a nontrivial scalar profile,
especially via the mechanism of spontaneous 
scalarization \cite{Damour:1993hw,Harada:1998ge,
Novak:1998rk,Sotani:2004rq,Cooney:2009rr,Arapoglu:2010rz,Orellana:2013gn,AparicioResco:2016xcm,Kase:2019dqc}. 
This includes Brans-Dicke theories \cite{Brans:1961sx} 
and $f(R)$ gravity \cite{Starobinsky:1980te}, where 
the latter corresponds to a particular class of the former 
with a scalar potential \cite{OHanlon:1972xqa,Chiba:2003ir}.
These nonminimally coupled theories belong to a subclass 
of non-shift-symmetric Horndeski theories. 

In shift-symmetric subclass of Horndeski theories 
where the field equations of motion are invariant under 
the shift $\phi \to \phi+c$, 
there is a no-hair argument of stars under several 
assumptions \cite{Lehebel:2017fag} 
analogous to BHs discussed in Ref.~\cite{Hui:2012qt}. 
The assumptions are as follows: 
\begin{enumerate}
\renewcommand{\theenumi}{\roman{enumi}}
\renewcommand{\labelenumi}{(\theenumi)}
\setlength{\itemsep}{0cm}
\item
\label{item1}
the scalar field and metrics are regular, static, and 
spherically symmetric with an asymptotically-flat 
spacetime geometry, 
\item
\label{item2}
a canonical kinetic term $X$ is present in the action, 
\item
\label{item3}
the action is analytic with regular coupling functions $G_{2,3,4,5}$.
\end{enumerate}
Under these hypotheses, we end up with a no-hair solution 
$\phi={\rm constant}$.

One way of breaking the assumption  
(\ref{item1}) is to postulate 
a scalar field of the form $\phi=qt+\psi(r)$, 
where $q$ is a nonvanishing constant, $t$ and $r$ are 
time and radial coordinates, respectively.  
Originally, this type of field configuration was considered to 
search for BH solutions
with a nontrivial profile of the scalar field in shift-symmetric Horndeski 
theories \cite{Babichev:2013cya}. 
The analysis was further extended to relativistic stars in DHOST theories \cite{Babichev:2016jom,Sakstein:2016oel,Kobayashi:2018xvr,Ogawa:2019gjc,Ikeda:2021skk}. 
In this paper, we do not consider such a time-dependent 
background scalar field and focus on the case $q=0$, i.e., the static scalar field.

If we break the assumption (\ref{item2}), i.e., 
no canonical kinetic term in the action, 
it is known that the quartic nonminimal derivative 
coupling $G_4 \supset \mu_4 X$ gives rise to 
a solution endowed with a nontrivial scalar profile
inside the star \cite{Cisterna:2015yla}. 
However, it was recognized that this solution with a
nontrivial scalar profile is plagued by an angular 
Laplacian instability of even-parity perturbations 
for large multipoles around the surface of star \cite{Kase:2020yjf,Kase:2021mix}.

Finally, one can break the above assumption 
(\ref{item3}) by introducing nonanalytic coupling functions. 
In the presence of $X$, the quintic-order coupling 
$G_5=-4 \alpha \ln |X|$, which is equivalent to the linear scalar-GB coupling $\alpha \phi R_{\rm GB}^2$, gives rise to 
NS solutions endowed with a nontrivial 
scalar profile \cite{Maselli:2016gxk}. 
This linear coupling can also accommodate a dilatonic coupling 
$\xi(\phi) \propto {\rm e}^{\mu \phi}$
in the limit $|\mu \phi| \ll 1$.
For the same scalar-GB coupling, there are also asymptotically-flat hairy 
BHs \cite{Sotiriou:2013qea,Sotiriou:2014pfa} consistent with 
all the linear stability conditions against odd- and 
even-parity perturbations \cite{Minamitsuji:2022mlv}. 
We note that BH solutions present for other nonanalytic 
functions in $G_{2,3,4}$ \cite{Babichev:2017guv} 
are either unstable around the horizon or 
asymptotically 
non-flat \cite{Creminelli:2020lxn,Minamitsuji:2022mlv}. 
For the linear scalar-GB coupling, it is not yet clear whether 
NS solutions with a nontrivial scalar profile satisfy 
all the stability conditions against odd- and 
even-parity perturbations. 
For this purpose, we can exploit conditions for the 
absence of ghost/Laplacian instabilities recently 
derived in full Horndeski theories \cite{Kase:2021mix} 
(see also Refs.~\cite{Kobayashi:2012kh,Kobayashi:2014wsa,Kase:2020qvz}).
NSs can have a nontrivial scalar profile in more general GB couplings $\alpha \xi(\phi) R_{\rm GB}^2$ with 
a canonical scalar kinetic term \cite{Pani:2011xm,Kleihaus:2014lba,Silva:2017uqg,Doneva:2017duq,Blazquez-Salcedo:2015ets,Olmo:2019flu}. 
This belongs to a subclass of non-shift-symmetric 
Horndeski theories. In this paper, for such general GB couplings, we will first study 
the background NS solutions and their linear stability both analytically and numerically.
We show that the strength of dimensionless coupling constant 
$|\alpha|$ has an upper bound to ensure the existence of NSs with a nontrivial scalar profile free from instabilities 
around the center of star. 
Besides $\alpha \xi(\phi) R_{\rm GB}^2$, 
we also incorporate regular 
coupling functions like $G_3 \supset \mu_3 X$,
$G_4 \supset \mu_4 X$, and $G_4 \supset \lambda_4 \phi$
as the representative cases, 
and study their effects on the existence and stability 
of solutions. Provided the coupling constants are 
in certain ranges, NS solutions with a nontrivial scalar 
profile consistent with all the linear stability conditions 
are still present in such combined theories. 

Moreover, there are also several other gravitational theories containing the GB term in the action. In so-called 4-dimensional-Einstein-GB (4DEGB) 
gravity \cite{Glavan:2019inb}, the contribution 
of the GB term 
in spacetime dimensions $D$ higher than 4 can be extracted 
by rescaling the GB coupling constant as $\alpha \to \alpha/(D-4)$.
If we perform a Kaluza-Klein reduction on a flat 
internal space whose volume is characterized by the 
scalar field $\phi$, the effective 
4DEGB theory after the rescaling of $\alpha$ 
belongs to 
a subclass of shift-symmetric Horndeski theories containing the 
linear GB coupling but without the canonical scalar kinetic term \cite{Lu:2020iav,Kobayashi:2020wqy} (see also Ref.~\cite{Fernandes:2020nbq,Hennigar:2020lsl} 
for a conformal regularization equivalent to the  Kaluza-Klein reduction).
It is known that NS solutions with a nontrivial scalar profile 
are present in the regularized 4DEGB theory \cite{Doneva:2020ped}, but we will show that 
they are plagued by a strong coupling problem and Laplacian
instability of even-parity perturbations. 
In theories given by the Lagrangian 
$R+F(R_{\rm GB}^2)$ \cite{Nojiri:2005jg,DeFelice:2007zq,Li:2007jm,DeFelice:2008wz,DeFelice:2009aj}, 
where $F$ is a positive power-law function of $R_{\rm GB}^2$, 
we will also show that a ghost instability and strong coupling 
at spatial infinity arise for NS solutions with a nontrivial 
scalar profile. 
In these 4DEGB and $F(R_{\rm GB}^2)$ theories, there are no canonical kinetic terms in the action,
which implies the unhealthy propagation of scalar field perturbations, as it also happens in derivative coupling theories without the canonical kinetic term \cite{Kase:2020yjf,Kase:2021mix}.

This paper is organized as follows.
In Sec.~\ref{scasec}, we present the linear stability 
conditions for relativistic stars on the
static and spherically symmetric background. 
In Sec.~\ref{GBsec}, we derive solutions expanded around the center of star and at spatial infinity in the presence of scalar-GB couplings $\alpha \xi(\phi) R_{\rm GB}^2$. 
We give a new theoretical bound on $\alpha$ for the existence of 
NS solutions with a nontrivial scalar profile free from instabilities and confirm it numerically 
for the linear scalar-GB coupling.
In Sec.~\ref{addsec}, we implement several regular coupling 
functions besides the scalar-GB coupling and explore the 
parameter space of coupling constants in which there are 
NS solutions with a nontrivial scalar profile without 
ghost or Laplacian instabilities.
In Sec.~\ref{4DEGBsec}, we show the existence of a strong coupling problem 
and Laplacian instability for 
NSs with a nontrivial scalar profile arising in 4DEGB gravity.
In Sec.~\ref{fGsec}, we prove that a non-vanishing scalar-field 
branch appearing in power-law $F(R_{\rm GB}^2)$ models is 
plagued by ghost and strong coupling problems 
at large distances. 
Sec.~\ref{consec} is devoted to conclusions.

\section{Background equations and linear 
stability conditions}
\label{scasec}

We study the existence and stability of NS solutions 
in Horndeski theories \cite{Horndeski,Def11,KYY,Charmousis:2011bf}, 
whose action is given by
\be
{\cal S}=\int {\rm d}^4 x \sqrt{-g}\,{\cal L}_H
+{\cal S}_m (g_{\mu \nu}, \Psi_m)\,,
\label{action}
\ee
where $g$ is a determinant of 
the metric tensor $g_{\mu \nu}$, and 
\ba
{\cal L}_H
&=&
G_2(\phi,X)-G_{3}(\phi,X)\square\phi 
+G_{4}(\phi,X)\, R +G_{4,X}(\phi,X)\left[ (\square \phi)^{2}
-(\nabla_{\mu}\nabla_{\nu} \phi)
(\nabla^{\mu}\nabla^{\nu} \phi) \right]
+G_{5}(\phi,X)G_{\mu \nu} \nabla^{\mu}\nabla^{\nu} \phi
\notag\\
&&
-\frac{1}{6}G_{5,X}(\phi,X)
\left[ (\square \phi )^{3}-3(\square \phi)\,
(\nabla_{\mu}\nabla_{\nu} \phi)
(\nabla^{\mu}\nabla^{\nu} \phi)
+2(\nabla^{\mu}\nabla_{\alpha} \phi)
(\nabla^{\alpha}\nabla_{\beta} \phi)
(\nabla^{\beta}\nabla_{\mu} \phi) \right]\,,
\label{LH}
\ea
where the coupling functions $G_{j}$ ($j=2,3,4,5$) depend on  
the scalar field $\phi$ and its kinetic term 
$X=-g^{\mu\nu}\nabla_{\mu}\phi\nabla_{\nu}\phi/2$, 
with the covariant derivative operator $\nabla_{\mu}$. 
We will use the notations  
$\square \phi \equiv \nabla^{\mu}\nabla_{\mu} \phi$ and 
$G_{j,\phi} \equiv \partial G_j/\partial \phi$, 
$G_{j,X} \equiv \partial G_j/\partial X$, 
$G_{j,\phi X} \equiv \partial^2 G_j/(\partial X \partial \phi)$, 
and so on. 
The scalar field $\phi$ and its derivatives are 
nonminimally coupled to the Ricci scalar $R$ and 
Einstein tensor $G_{\mu \nu}$ 
through the couplings $G_4$ and $G_5$, respectively.
For the matter fields $\Psi_m$ inside NSs, we consider a perfect fluid given by the energy-momentum tensor 
\be
T_{\mu \nu}=\left( \rho+P \right) u_{\mu} u_{\nu}+P g_{\mu \nu}\,,
\ee
where $\rho$ and $P$ are the density and pressure, respectively, 
and $u_{\mu}$ is the four-velocity of the fluid satisfying the normalization relation
$u_{\mu}u^{\mu}=-1$.
Assuming that the perfect fluid is minimally coupled to gravity, 
it obeys the continuity equation 
\be
\nabla^{\mu} T_{\mu \nu}=0\,.
\label{Tcon}
\ee
In terms of the action approach, the perfect fluid can be described by a Schutz-Sorkin action \cite{Schutz:1977df,Brown:1992kc,DeFelice:2009bx}.

\subsection{Background equations of motion}

A static and spherically symmetric background is described 
by the line element
\be
\rd s^2=-f(r) \rd t^{2} +h^{-1}(r) \rd r^{2}
+ r^{2} \left(\rd \theta^{2}
+\sin^{2}\theta\,\rd\varphi^{2} 
\right)\,,
\label{BGmetric}
\ee
where $f(r)$ and $h(r)$ are functions of the radial
coordinate $r$. 
On this background, we consider the scalar field
that depends only on the radial coordinate
\be
\phi=\phi(r)\,,
\ee
together with the four-velocity of the fluid $u^{\mu}=[f(r)^{-1/2},0,0,0]$. 
Then, the mixed energy-momentum tensor $T^{\mu}{}_{\nu}$ has 
the following diagonal components
\be
T^{\mu}{}_{\nu}={\rm diag} \left[ 
-\rho(r), P(r), P(r), P(r) \right]\,,
\ee
where $\rho$ and $P$ are functions of $r$ alone. 
The continuity Eq.~(\ref{Tcon}) gives 
\be
P'+\frac{f'}{2f} \left( \rho+P \right)=0\,,
\label{mattereq}
\ee
where a prime represents the derivative with respect to $r$.

The (00), (11), (22) components of gravitational 
field equations of motion are 
\ba
& &
\left(A_1+\frac{A_2}{r}+\frac{A_3}{r^2}\right)\phi''
+\left(\frac{\phi'}{2h}A_1+\frac{A_4}{r}+\frac{A_5}{r^2}\right)h'
+A_6+\frac{A_7}{r}+\frac{A_8}{r^2}=\rho
\,,\label{back1}\\
& &
-\left(\frac{\phi'}{2h}A_1+\frac{A_4}{r}+\frac{A_5}{r^2}\right) \frac{hf'}{f}
+A_9-\frac{2\phi'}{r}A_1-\frac{1}{r^2}\left[\frac{\phi'}{2h}A_2+(h-1)A_4\right]=P
\,,\label{back2}\\
& &
\left[\left\{A_2+\frac{(2h-1)\phi'A_3+2hA_5}{h\phi' r}\right\}
\frac{f'}{4f}+A_1+\frac{A_2}{2r}\right]\phi''
+\frac{1}{4f}\left(2hA_4-\phi'A_2+\frac{2hA_5-\phi'A_3}{r}\right)\left(f''-\frac{f'^2}{2f}\right) \nonumber \\
&&
+\left[A_4+\frac{2h(2h+1)A_5-\phi'A_3}{2h^2r}\right]\frac{f'h'}{4f}
+\left(\frac{A_7}{4}+\frac{A_{10}}{r}\right)\frac{f'}{f}
+\left(\frac{\phi'}{h}A_1+\frac{A_4}{r}\right)\frac{h'}{2}+A_6+\frac{A_7}{2r}
=-P\,,\label{back3}
\ea
where the coefficients $A_1$-$A_{10}$ are given in Appendix \ref{AppA}.
The scalar-field equation of motion is expressed in the form 
\be
\frac{1}{r^2} \sqrt{\frac{h}{f}} 
\left( r^2 \sqrt{\frac{f}{h}} J^r 
\right)'+{\cal P}_{\phi}=0\,,
\label{Ephi2}
\ee
with
\ba
J^r &=&
h \phi' \biggl[
G_{2,X}-\left( \frac{2}{r}+\frac{f'}{2f} 
\right) h \phi' G_{3,X}+2 \left( \frac{1-h}{r^2}
-\frac{h f'}{rf} \right) G_{4,X}+2h \phi'^2 
\left( \frac{h}{r^2}+\frac{hf'}{rf} \right) G_{4,XX} 
\nonumber \\
&& \qquad 
-\frac{f'}{2r^2f} (1-3h)h \phi' G_{5,X}
-\frac{f' h^3 \phi'^3}{2r^2 f}G_{5,XX} \biggr]\,,
\label{Jrc}\\
{\cal P}_{\phi} &=& G_{2,\phi}+\lambda_1 G_{3,\phi}+\lambda_2 G_{3,\phi \phi}
+\lambda_3 G_{3,\phi X}+\lambda_4 G_{4,\phi}+\lambda_5 G_{4,\phi X}
+\lambda_6 G_{4,\phi \phi X}+\lambda_7 G_{4,\phi XX} \nonumber \\
&&
+\lambda_8 G_{5,\phi}
+\lambda_9 G_{5,\phi \phi}+\lambda_{10} G_{5,\phi X}+
\lambda_{11} G_{5,\phi \phi X}+\lambda_{12}G_{5,\phi X X}\,,
\label{Pphide}
\ea
where $\lambda_1$--$\lambda_{12}$ are presented 
in Appendix~\ref{AppA}. 
This equation also follows by combining 
Eqs.~(\ref{back1})-(\ref{back3}).
Note that, in shift-symmetric Horndeski theories where the coupling functions 
$G_j$ contain the $X$ dependence alone, we have ${\cal P}_{\phi}=0$.
In this case, Eq.~(\ref{Ephi2}) gives the solution 
$J^r=(Q/r^2) \sqrt{h/f}$, where $Q$ is a constant.

\subsection{Linear stability conditions}

To study the linear stability of NS solutions, we consider metric perturbations $h_{\mu \nu}$ on top of the background (\ref{BGmetric}) besides perturbations of the scalar field 
and perfect fluid. Expanding perturbations on the 
background (\ref{BGmetric}) in terms of the spherical harmonics
of the unit two-sphere $Y_{l m} (\theta, \varphi)$,
one can decompose them into the two different sectors
depending on the parity under the rotation 
along two-dimensional 
sphere \cite{Regge:1957td,Zerilli:1970se}.
The odd- and even-parity perturbations have 
the parities $(-1)^{l+1}$ and $(-1)^l$, respectively.
Any scalar perturbation has the even mode alone, whereas vector and tensor perturbations contain both 
odd and even modes. 
The decomposition of perturbations of metrics, scalar field, 
and perfect fluids into the odd- and even-parity modes was 
addressed in Ref.~\cite{DeFelice:2011ka,Motohashi:2011pw,Kobayashi:2012kh,Kobayashi:2014wsa,Kase:2020qvz,Kase:2021mix}. 
In the presence of perfect fluids the stability conditions against odd- and even-parity perturbations were already derived 
in Ref.~\cite{Kase:2021mix}, so we briefly 
summarize them in the following. 

In the odd-parity sector, there is a dynamical 
perturbation $\chi$ arising from the gravity sector 
besides a nondynamical perturbation $\delta j$ related to 
the $(\theta, \varphi)$ components of fluid four 
velocity \cite{Kase:2021mix}. 
In the limit of large frequencies and multipoles $l$, 
the no-ghost condition for the dynamical field $\chi$ 
translates to 
\be
{\cal G} \equiv 2 G_4+2 h\phi'^2G_{4,X}-h\phi'^2 
\left( G_{5,\phi}+\frac{f' h\phi' G_{5,X}}{2f} \right)>0\,.
\label{cGdef}
\ee
Under this condition, the Laplacian instability along 
the radial and angular directions can be avoided for
\ba
{\cal H}&\equiv&2 G_4+2 h\phi'^2G_{4,X}-h\phi'^2G_{5,\phi}
-\frac{h^2 \phi'^3 G_{5,X}}{r}>0
\,,\label{cHdef}\\
{\cal F}&\equiv&2 G_4+h\phi'^2G_{5,\phi}-h\phi'^2 
\left( \frac12 h' \phi'+h \phi'' \right) G_{5,X}>0\,,
\label{cFdef}
\ea
under which the squared propagation speeds 
$c_r^2={\cal G}/{\cal F}$ and 
$c_\Omega^2={\cal G}/{\cal H}$ are positive.
The expressions (\ref{cGdef})-(\ref{cFdef}) coincide with those 
originally derived in Ref.~\cite{Kobayashi:2012kh} 
in the absence of the perfect fluid. 

In the even-parity sector, there are three dynamical perturbations:
matter perturbation $\delta \rho$, 
gravitational perturbations $\psi$, and 
scalar-field perturbation $\delta \phi$. 
We focus on the linear stability conditions of high radial and angular momentum modes. With the condition (\ref{cHdef}), 
there are no ghosts for even-parity perturbations if 
\ba
& &
\rho+P>0\,,\label{nogoma}\\ 
& &
{\cal K} \equiv (2{\cal P}_1-{\cal F})
h \mu^2 -2{\cal H} ^2 r^4 (\rho+P)>0\,,
\label{Kcon}
\ea
where
\be
{\cal P}_1 \equiv \frac{h \mu}{2fr^2 {\cal H}^2} 
\left( 
\frac{fr^4 {\cal H}^4}{\mu^2 h} \right)'\,,\qquad
\mu \equiv \frac{2(\phi' a_1+r\sqrt{fh}{\cal H})}{\sqrt{fh}}\,.
\label{defP1}
\ee
The definition of $a_1$ is given in Appendix \ref{AppB}.

The radial propagation speed squared of $\psi$ 
is given by $c_{r2}^2={\cal G}/{\cal F}$, which is the same as 
that of $\chi$. 
Thus the gravitational perturbations in the 
odd- and even-parity sectors propagate in the same manner 
along the radial direction.
The radial Laplacian instabilities of $\delta \rho$ 
and $\delta \phi$ can be avoided for 
\ba
& &
c_{m}^2 \equiv \frac{n \rho_{,n n}}{\rho_{,n}}>0\,,
\label{cmcon} \\
& &
c_{r3}^2 \equiv
\frac{2\phi'[ 4r^2 (fh)^{3/2} {\cal H} c_4 
(2\phi' a_1+r\sqrt{fh}\,{\cal H})
-2a_1^2 f^{3/2} \sqrt{h} 
\phi' {\cal G} 
+( a_1 f'+2 c_2 f ) r^2 fh 
{\cal H}^2]}{f^{5/2} \sqrt{h}\,{\cal K}}>0
\,,\label{cr3}
\ea
where $c_2$ and $c_4$ are given in Appendix \ref{AppB}.
The conditions (\ref{nogoma}) and (\ref{cmcon}) correspond to 
those in the perfect fluid sector. We will consider the fluid
equation of state (EOS) satisfying these inequalities. 
We note that $c_{r3}^2$ corresponds to
the propagation speed squared of scalar-field 
perturbation $\delta \phi$.
  
Along the angular direction, the perfect fluid in the 
even-parity sector has the propagation speed squared same as 
$c_m^2=n \rho_{,n n}/\rho_{,n}$. 
The angular Laplacian instabilities for $\psi$ and $\delta \phi$ 
are absent if
\be
c_{\Omega\pm}^2=-B_1\pm\sqrt{B_1^2-B_2}>0\,,
\label{cosq}
\ee
where 
\ba
\hspace{-0.9cm}
&&
B_1 \equiv
\frac {r^3\sqrt {f h} {\cal H} 
[4 h \beta_0 \beta_1+\beta_2-4 \phi' a_1 \beta_3
+r \sqrt{f^3 h}{\cal G}{\cal H} (\rho+P)]
-2 fh {\cal G}  [ r \sqrt{fh}( 2 {\cal P}_1-{\cal F}){\cal H}  
( \beta_0+\phi' a_1 )+2\phi'^2a_1^{2}{\cal P}_1 ] }
{4f h ( 2 {\cal P}_1-{\cal F} ){\cal H}\beta_0^2
-2 r^4 f^2 h  {\cal H}^3 (\rho+P)},
\label{B1def}\\
\hspace{-0.9cm}
&&
B_2 \equiv
-2r^2{\frac {r^2h \beta_1 ( 2 fh {\cal F} {\cal G}\beta_0 
+r^2\beta_2) -{r}^{4}\beta_2 \beta_3
-fh{\cal F} {\cal G}  ( \phi' fh {\cal F} {\cal G}a_1 +2 r^3 \sqrt{fh} {\cal H} \beta_3 )}
{2fh \phi' a_1 (2 {\cal P}_1-{\cal F}){\cal F} \beta_0^{2}
-r^4 f^2 h \phi' a_1 {\cal F} {\cal H}^2 (\rho+P)} }\,.
\label{B2def}
\ea
The explicit forms of $\beta_0, \beta_1, \beta_2, \beta_3$ are 
presented in Appendix \ref{AppB}. 
The stability conditions (\ref{cosq}) are satisfied if 
\be
B_1^2 \geq B_2>0\,,\quad {\rm and} \quad
B_1<0\,.
\label{B12con}
\ee
In summary, we require that NS solutions with 
a nontrivial scalar profile should satisfy 
the inequalities (\ref{cGdef}), (\ref{cHdef}), (\ref{cFdef}), 
(\ref{Kcon}), (\ref{cr3}), and (\ref{cosq}) besides 
the fluid stability conditions (\ref{nogoma}) and (\ref{cmcon}).

\section{Scalar-Gauss-Bonnet couplings}
\label{GBsec}

Let us first consider the Einstein-scalar-GB theory
given by the action 
\be
{\cal S}=\int {\rm d}^4 x \sqrt{-g} 
\left[ \frac{\Mpl^2}{2}R
+\eta X+\alpha \xi(\phi)R_{\rm GB}^2 \right]\,,
\label{GBaction}
\ee
where $\eta$ is a constant,
$\Mpl$ is the reduced Planck mass, 
$\alpha$ is a dimensionless coupling,
$\xi$ is a function of $\phi$, and 
\be
\label{def_gb}
R_{\rm GB}^2
\equiv R^2-4R_{\alpha\beta}R^{\alpha\beta}
+R_{\alpha\beta\mu\nu}R^{\alpha\beta\mu\nu}\,,
\ee
with $R_{\alpha\beta}$ and $R_{\alpha\beta\mu\nu}$ being 
the Ricci and Riemann tensors respectively.
The action (\ref{GBaction}) belongs to a subclass of 
Horndeski theories with the coupling functions
\ba
& &
G_2=
\eta X+8 \alpha \xi^{(4)}(\phi) X^2 (3-\ln |X|)\,,\qquad 
G_3=4\alpha \xi^{(3)}(\phi) X (7-3\ln |X|)\,,\nonumber \\
& &
G_4=\frac{\Mpl^2}{2}+4\alpha\xi^{(2)}(\phi) X (2-\ln |X|)\,,\qquad
G_5=-4\alpha\xi^{(1)}(\phi) \ln |X|\,,
\label{Ggauss}
\ea
where $\xi^{(n)}(\phi) \equiv {\rm d}^n \xi(\phi)/{\rm d} \phi^n$.

The background Eqs.~(\ref{back1}), (\ref{back2}), 
and (\ref{Ephi2}) reduce, respectively, to 
\ba
& &
h' = \frac{2(1-h)[\Mpl^2-8\alpha h (\phi''\xi_{,\phi}
+\phi'^2 \xi_{,\phi \phi})]-r^2 (2\rho+\eta h \phi'^2)}
{2 \Mpl^2 r+8 \alpha (1-3h)\phi' \xi_{,\phi}}\,,
\label{GBb1}\\
& &
f' = \frac{f [2(1-h) \Mpl^2+r^2 
(2P+\eta h \phi'^2)]}
{h[2 \Mpl^2 r+8 \alpha (1-3h)\phi' \xi_{,\phi}]}\,,
\label{GBb2}\\
& &
\eta h \phi''+\frac{\eta (f' h r+f h' r+4fh)}{2rf}\phi'
-\frac{2\alpha \xi_{,\phi} [h (1-h)(2f f''-f'^2)+f f' h' (1-3h)]}{r^2 f^2}=0\,.
\label{GBb3}
\ea
The linear stability conditions 
(\ref{cGdef})-(\ref{cFdef}) 
in the odd-parity sector translate to 
\ba
{\cal G} &= &\Mpl^2-\frac{4\alpha 
\xi_{,\phi} \phi'f' h}{f}>0\,,\\
{\cal H} &= &\Mpl^2-\frac{8\alpha 
\xi_{,\phi} \phi' h}{r}>0\,,\\
{\cal F} &=& \Mpl^2-4 \alpha \left( 
2h \xi_{,\phi \phi} \phi'^2  
+2h \xi_{,\phi} \phi''+h' \xi_{,\phi} \phi' 
\right)>0\,.
\ea
In the limit of a small GB coupling $\alpha \to 0$, 
all of ${\cal G}$, ${\cal H}$, and ${\cal F}$ 
approach $\Mpl^2$,
so the stability against 
odd-parity perturbations is ensured. 
As we will see later in 
Sec.~\ref{censec}, the leading-order term of
$\phi'$ around $r=0$ is proportional to $r$,
so ${\cal H}$ does not diverge at the center of star.

On using Eqs.~(\ref{GBb1}) and (\ref{GBb2}) to 
eliminate $\rho$ and $P$, the no-ghost condition (\ref{Kcon}) in the even-parity 
sector yields
\ba
{\cal K} &=& 2\eta h \phi'^2 r^2 (\Mpl^2 r-8 \alpha 
 h  \xi_{,\phi} \phi')^2 \nonumber\\
& &
+64 \alpha^2 h(h-1) \phi'^2 \xi_{,\phi}^2  
[\Mpl^2 (2 r h'-h+1)-4\alpha \xi_{,\phi} \phi' 
h'(1+3h)+8\alpha h (h-1) (\xi_{,\phi}\phi''+
\xi_{,\phi \phi} \phi'^2)]>0\,.
\ea
In the limit $\alpha \to 0$, we have 
${\cal K} \to 2\eta h \phi'^2 \Mpl^4 r^4$ and hence the ghost can be avoided for 
\be
\eta>0\,.
\ee
In the following, we will focus on the case in which $\eta$ is positive.

The condition (\ref{cr3}) for the absence of
the Laplacian instability along the radial direction reduces to
\be
c_{r3}^2=\frac{2h \phi'^2 [\eta f r^2 (\Mpl^2 r
-8 \alpha h \xi_{,\phi} \phi')^2-32 \alpha^2 
(h-1) \xi_{,\phi}^2 \{ \Mpl^2 [f(h-1)-2rf' h]
+4\alpha \xi_{,\phi}\phi' f'h(1+3h)\}]}{f{\cal K}}>0\,.
\label{cr3a}
\ee
Expanding $c_{r3}^2$ around $\alpha=0$, we obtain
\be
c_{r3}^2=1-\frac{64 \xi_{,\phi}^2 (1-h) 
(f' h -f h')}{\eta f \Mpl^2 r^3} \alpha^2
+{\cal O}(\alpha^3)\,,
\label{cr3GB}
\ee
and hence $c_{r3}^2 \to 1$ as $\alpha \to 0$.

The squared angular propagation speeds of even-parity perturbations 
are complicated, but expanding 
$c_{\Omega,\pm}^2$ around $\alpha=0$ leads to 
\ba
c_{\Omega,\pm}^2=1+\frac{\xi_{,\phi}
[2\sqrt{\eta}fh (2f-rf')r\phi' \pm \sqrt{2{\cal B}}]}
{\sqrt{\eta} f^2 \Mpl^2 r^2}|\alpha|
+{\cal O}(\alpha^2)\,,
\ea
where
\be
{\cal B} \equiv \Mpl^2 [f'^2 h r^2 + 2f^2(r h' - 2h + 2) 
- rf (2f'' hr + rf'h' - 2f'h)]^2 
+ 2\eta f^2 h^2 (r f' - 2f)^2 r^2 \phi'^2\,.
\ee
Under the no-ghost condition $\eta>0$, ${\cal B}$ 
is positive and hence $c_{\Omega,\pm}^2$ are the real values. 
In the limit $\alpha \to 0$, $c_{\Omega,\pm}^2$ 
approach 1.

{}From the above discussion, all the linear stability conditions
should be consistently satisfied for $\eta>0$ and $|\alpha| \ll 1$. 
We note, however, that the quantities like 
${\cal H}$, $c_{r3}^2$, $c_{\Omega,\pm}^2$ contain 
positive power-law terms of $r$ in the denominators. 
To show the finiteness of these quantities at $r=0$, 
we derive the solutions to $f$, $h$, $\phi$ 
expanded around $r=0$ in Sec.~\ref{censec}. 
For the existence of hairy stars free from instabilities, 
we then put a limit on the coupling constant $\alpha$.

\subsection{Solutions expanded around $r=0$ and their stability}
\label{censec}

Around the center of star, we impose 
the regular boundary conditions 
$f(0)=f_c$, $h(0)=1$, $\phi(0)=\phi_c$, 
$\rho(0)=\rho_c$, $P(0)=P_c$ and 
$f'(0)=h'(0)=\phi'(0)=\rho'(0)=P'(0)=0$. 
Then, around $r=0$, the scalar field is 
expanded as 
\be
\phi=\phi_c+\phi_2 r^2
+{\cal O}(r^3)\,,
\label{phir=0}
\ee
likewise for $f$, $h$, $\rho$, and $P$.
We can also expand the coupling function 
$\xi(\phi)$ (and its $\phi$ derivatives), as
\be
\xi(\phi)=\xi(\phi_c)+\sum_{n 
\geq 1} \xi^{(n)}(\phi_c)\frac{(\phi-\phi_c)^n}
{n!}\,.
\label{xiphi}
\ee
On using the background 
Eqs.~(\ref{GBb1})-(\ref{GBb3}) with Eq.~(\ref{mattereq}), the quantity $\phi_2$ in Eq.~(\ref{phir=0}) obeys 
the following algebraic equation 
\be
Y(x) \equiv \frac{9x (1-16x)^3}{2[1+3w_c(1-16x)]} 
=\kappa \alpha^2\,,
\label{Ykappa}
\ee
where 
\be
x \equiv \frac{\alpha \xi_{,\phi}(\phi_c)}{\Mpl^2}\phi_2\,,
\qquad 
\kappa \equiv \frac{\rho_c^2 \xi_{,\phi}^2(\phi_c)}
{\eta \Mpl^6}\,,
\qquad
w_c=\frac{P_c}{\rho_c}\,.
\label{xdef}
\ee
We assume that the EOS parameter $w_c$ is 
in the range $w_c>0$.
In the small-coupling limit $|\alpha| \ll 1$, we have 
$9x/[2(1+3w_c)] \simeq \kappa \alpha^2$ and hence
\be
\phi_2=\frac{2 \xi_{,\phi}(\phi_c)
\rho_c^2 (1+3w_c)}{9 \eta \Mpl^4}\alpha
+{\cal O}(\alpha^3)\,.
\label{phi2}
\ee
Substituting the leading-order solution of Eq.~(\ref{phi2}) 
into the definition of $x$ in Eq.~(\ref{xdef}),  
we have 
\be
x>0\,,
\label{xcon1}
\ee
under the condition $\eta>0$. 
By the end of this section, 
we will {\it not} exploit the small $\alpha$ expansion 
to discuss the linear stability of NSs.

The solutions to $f$, $h$, $P$ expanded around $r=0$  
are given, respectively, by 
\ba
f &=& f_c+\frac{f_c [ \Mpl^2 \rho_c (1+3w_c)
-48 \alpha \xi_{,\phi}(\phi_c) \rho_c w_c  \phi_2]}
{6 [\Mpl^2-16 \alpha \xi_{,\phi}(\phi_c)\phi_2]^2}r^2
+{\cal O}(r^3)\,,\label{fr=0} \\
h &=& 1-\frac{\rho_c}{3[\Mpl^2-16 \alpha \xi_{,\phi}(\phi_c)\phi_2]}r^2+{\cal O}(r^3)\,,
\label{hr=0}\\
P &=& P_c-\frac{\rho_c (1+w_c)[ \Mpl^2 \rho_c (1+3w_c) 
-48 \alpha \xi_{,\phi}(\phi_c) \rho_c w_c \phi_2]}
{12 [\Mpl^2-16 \alpha \xi_{,\phi}(\phi_c)\phi_2]^2}r^2
+{\cal O}(r^3)\,.
\label{Pr=0}
\ea
The scalar field is of the form (\ref{phir=0}) with $\phi_2$ satisfying the relation (\ref{Ykappa}). 
On using these solutions, the quantities ${\cal G}$, ${\cal H}$, and ${\cal F}$ at $r=0$ reduce to 
\be
{\cal G}(r=0)=\Mpl^2\,,\qquad 
{\cal H}(r=0)={\cal F}(r=0)
=\Mpl^2 \left( 1-16x \right)\,,
\label{GHF}
\ee
where ${\cal H}$ is finite at $r=0$ due to the property that
$\phi'=2 \phi_2 r$ at leading order.
The squared propagation speeds of odd-parity perturbations 
along the radial and angular directions are given by 
\be
c_r^2(r=0)=c_{\Omega}^2(r=0)=\frac{1}{1-16x}\,.
\label{crOr=0}
\ee
Then, there are neither ghost nor Laplacian instabilities 
in the odd-parity sector if
\be
0<x<\frac{1}{16}\,,
\label{xcon2}
\ee
where we have also taken into account the condition (\ref{xcon1}).
Both $c_r^2(r=0)$ and $c_{\Omega}^2(r=0)$ 
are larger than 1.
In the limit that $\alpha \to 0$, we have $x \to 0$ and 
hence $c_r^2(r=0)=c_{\Omega}^2(r=0) \to 1$ 
as expected. 
In the subject of NSs, ``superluminality" is sometimes argued as 
a sign of ``acausality.'' 
We emphasize that the propagation speed 
of scalar-field perturbations just fixes a causal boundary of the scalar field at each position, 
and the superluminal speeds do not mean acausality.

In the even-parity sector, the leading-order term of ${\cal K}$ expanded around $r=0$ is proportional to $r^6$, i.e., 
\be
{\cal K}=\frac{16 \rho_c^2 \Mpl^2 
x[1+48x +3w_c(1-16x)]}{9(1-16x)}r^6+{\cal O}(r^{7})\,,
\label{Kest}
\ee
where we used Eq.~(\ref{Ykappa}) to eliminate $\kappa \alpha^2$. 
Under the condition (\ref{xcon2}), the coefficient of 
$r^6$ in Eq.~(\ref{Kest}) is positive.
The squared radial and angular propagation speeds 
of $\delta \phi$ at $r=0$ are given by 
\be
c_{r3}^2(r=0)=c_{\Omega -}^2(r=0)
=\frac{(1+3w_c)(1-64 x)+2304 w_c x^2}
{(1-16x)[1+48 x+3w_c (1-16x)]}\,, 
\label{cr3GBa}
\ee
whereas $c_{\Omega +}^2(r=0)$ is equivalent to Eq.~(\ref{crOr=0}).
Under the condition (\ref{xcon2}), the absence of 
Laplacian instability of $\delta \phi$ requires a positivity 
of the numerator of Eq.~(\ref{cr3GBa}), so that 
\be
0<x<x_m \equiv \frac{2(1+3w_c)-\sqrt{(1+3w_c)(4+3w_c)}}
{144 w_c}\,.
\label{xrange}
\ee
For $0<w_c<\infty$ we have $1/64<x_m<1/48$, so 
$x_m$ is smaller than $1/16$.
The function $Y(x)$ in Eq.~(\ref{Ykappa}) has a maximum value 
$Y(x_m)$ at $x=x_m$.
Provided that $0<\kappa \alpha^2<Y(x_m)$, there are solutions to Eq.~(\ref{Ykappa}). 
This gives an upper bound on $|\alpha|$, as 
\be
|\alpha|<\frac{\sqrt{\eta}\Mpl^3}{\rho_c \xi_{,\phi}(\phi_c)} 
\frac{\sqrt{6}(2\sqrt{1+3w_c}-\sqrt{4+3w_c})^{1/2}
[\sqrt{(1+3w_c)(4+3w_c)}-2+3w_c]^{3/2}}
{216 w_c^2 (\sqrt{1+3w_c}+\sqrt{4+3w_c})^{1/2}}\,.
\label{alphacon}
\ee
Among the two solutions of $\kappa \alpha^2=Y(x)$, 
one of them ($x=x_1$) exists in $0<x_1<x_m$, 
while the other ($x=x_2$) is in the region $x_m<x_2<1/16$. 
The former is in the region (\ref{xrange}) satisfying the 
condition of Laplacian stability.
In summary, as long as $\alpha$ is in the range (\ref{alphacon}), 
there is a solution $x=x_1$ consistent will all the linear 
stability conditions at $r=0$. 
The criterion (\ref{alphacon}) is also valid for nonrelativistic stars
with $w_c \ll 1$. In this case, the scalar-GB coupling is constrained 
to be
\be
|\alpha|<\frac{9\sqrt{6\eta}\Mpl^3}{128\rho_c \xi_{,\phi}(\phi_c)}\,,
\quad {\rm for} \quad w_c \to 0\,,
\label{alphacon2}
\ee
which can be applied to nonrelativistic objects such as Sun and Earth.

\begin{figure}[ht]
\begin{center}
\includegraphics[height=3.2in,width=3.5in]{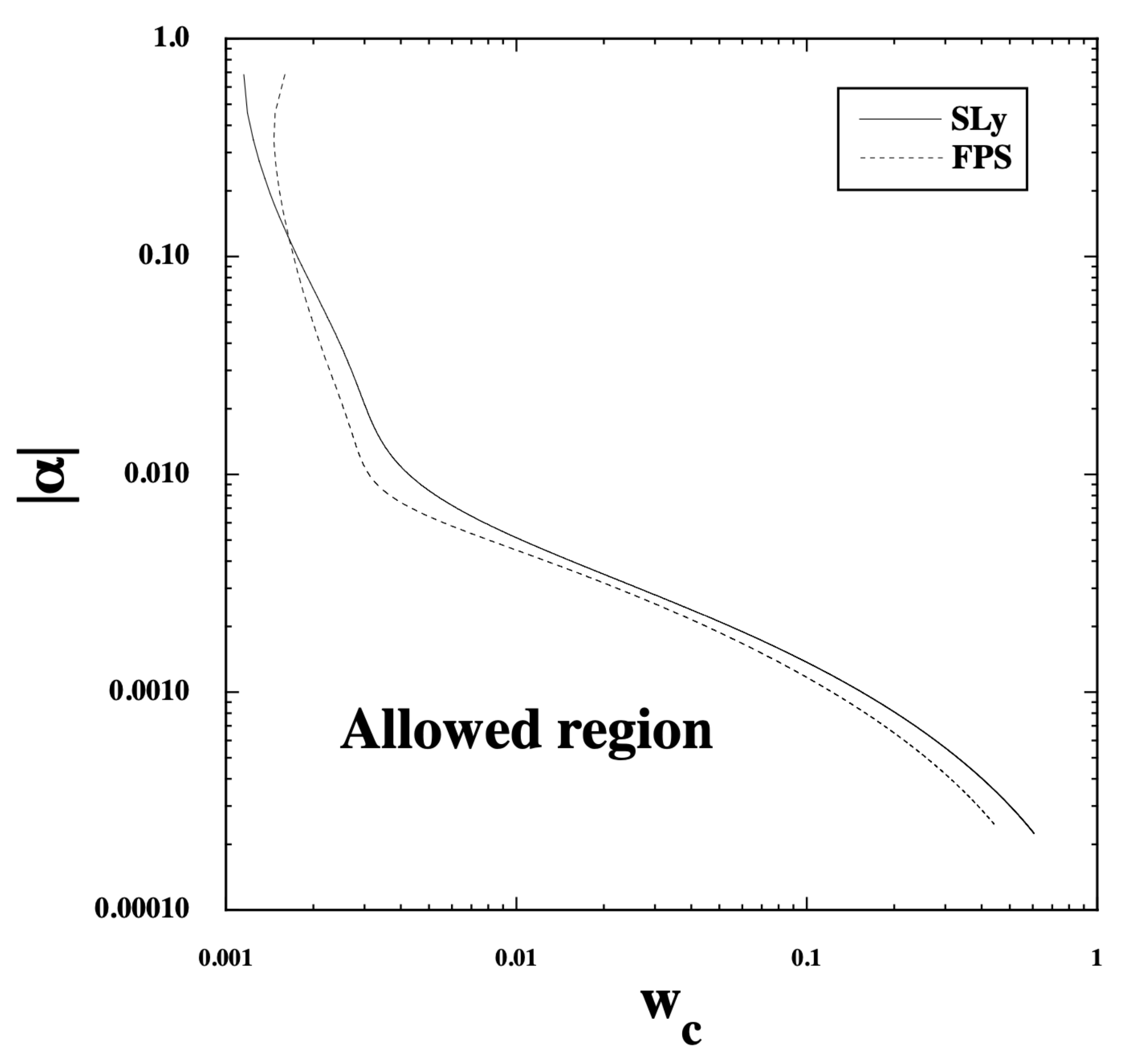}
\end{center}\vspace{-0.5cm}
\caption{Maximum value of $|\alpha|$ versus $w_c$  
constrained by the bound (\ref{alpocon}) for the linear GB coupling 
($n=1$) with $\eta=1$. The solid and dashed lines correspond to 
the upper limits $|\alpha_{\rm max}|$ for SLy and FPS EOSs, 
respectively, with the central density in the range 
$10^{-2} \rho_0 \le \rho_c \le 20 \rho_0$.
\label{fig1}
}
\end{figure}

Let us consider the power-law 
scalar-GB coupling
\be
\xi(\phi)=\Mpl^{2-n} r_0^2 \phi^n\,,
\label{powerGB}
\ee
where 
\be
r_0=\sqrt{\frac{8\pi \Mpl^2}{\rho_0}}=89.664~{\rm km}\,,
\qquad \rho_0=m_n n_0=1.6749 \times 
10^{14}~{\rm g} \cdot {\rm cm}^{-3}\,.
\ee
Here, $m_n=1.6749 \times 10^{-24}$~g is the neutron 
mass and $n_0=0.1~(\rm fm)^{-3}$ is the typical 
density of NSs.
For this coupling, the bound (\ref{alphacon}) yields 
\be
|\alpha|<|\alpha_{\rm max}| \equiv \frac{\sqrt{\eta}}{8 \pi n} 
\frac{\rho_0}{\rho_c}
\left( \frac{\phi_c}{\Mpl} \right)^{1-n} 
\frac{\sqrt{6}(2\sqrt{1+3w_c}-\sqrt{4+3w_c})^{1/2}
[\sqrt{(1+3w_c)(4+3w_c)}-2+3w_c]^{3/2}}
{216 w_c^2 (\sqrt{1+3w_c}+\sqrt{4+3w_c})^{1/2}}\,.
\label{alpocon}
\ee
For the linear GB coupling ($n=1$), the central density 
$\rho_c$ and EOS parameter $w_c$ determine the upper 
limit of $|\alpha|$, without having the dependence of $\phi_c$.
In Fig.~\ref{fig1}, we plot $|\alpha_{\rm max}|$
as a function of $w_c$ for SLy (solid) and FPS (dashed) 
EOSs \cite{Haensel:2004nu} in the range $10^{-2} \rho_0 \le \rho_c \le 20 \rho_0$.
As $\rho_c$ increases, $w_c$ grows from nonrelativistic 
values of order $10^{-3}$ to relativistic values of order $10^{-1}$. 
We find that $|\alpha_{\rm max}|$ is a decreasing function 
of $w_c$. For SLy EOS, the maximum mass of NS is reached 
around the central density $\rho_c \approx 15 \rho_0$ 
with the EOS parameter $w_c \approx 0.5$. 
To realize such a maximum mass of NS, the scalar-GB coupling 
is constrained to be 
\be
|\alpha|<3 \times 10^{-4}\,.
\label{alphabo}
\ee
For FPS EOS, we also obtain a similar upper bound. 
The existence of NS solutions with the central density 
$\rho_c \gtrsim 15 \rho_0$ gives the value of $|\alpha_{\rm max}|$ 
even smaller than $3 \times 10^{-4}$. 

The dilatonic GB coupling $\alpha_{\rm GB}$ in the regime 
$\phi/\Mpl \ll 1$, which was discussed in 
Refs.~\cite{Saffer:2021gak,Lyu:2022gdr} 
with the unit $M_{\rm pl}=1/\sqrt{8 \pi}$, 
is related to our linear scalar-GB coupling $\alpha$ by 
$\alpha_{\rm GB}=\alpha \Mpl r_0^2 =\alpha r_0^2/\sqrt{8 \pi}$. 
Then, the theoretical bound (\ref{alphabo}) translates to 
\be
\sqrt{|\alpha_{\rm GB}|}<0.7~{\rm km}\,.
\ee
This is tighter than the typical observational bounds 
$\sqrt{\alpha_{\rm GB}}<{\cal O}(1)~{\rm km}$ derived from 
the GW measurements of NS-NS, BH-NS, and BH-BH 
binaries, see Table I of Ref.~\cite{Lyu:2022gdr}. 
Thus, the existence of NS solutions consistent with 
the linear stability conditions gives a new theoretical 
upper bound on the dilatonic GB coupling.

\subsection{Solutions expanded at spatial infinity 
and their stability}
\label{infsec}

The surface of star is defined by the radius $r_s$ 
at which the fluid pressure $P$ vanishes. 
Outside the star ($r>r_s$), we have $\rho=0=P$ 
in the background Eqs.~(\ref{back1})-(\ref{Ephi2}).
Imposing the asymptotic flatness at spatial infinity, 
we can expand $f$, $h$, and $\phi$ in the forms 
$f=1+\sum_{i=1}\hat{f}_i/r^i$, 
$h=1+\sum_{i=1}\hat{h}_i/r^i$, and 
$\phi=\phi_0+\sum_{i=1}\hat{\phi}_i/r^i$. 
We also use the expansions of $\xi(\phi)$ and 
its $\phi$ derivatives analogous to Eq.~(\ref{xiphi}), with the 
replacement $\phi_c \to \phi_0$.
The large-distance solutions consistent with 
the background Eqs.~(\ref{GBb1})-(\ref{GBb3}) are 
given by 
\ba
f &=& 1-\frac{2M}{r}+\frac{\eta M \hat{\phi}_1^2}
{6 \Mpl^2 r^3}+\frac{M \hat{\phi}_1 [\eta M \hat{\phi}_1
+24 \alpha \xi_{,\phi}(\phi_0)]}{3 \Mpl^2 r^4}
+{\cal O}(r^{-5})\,,
\label{finf} \\
h &=& 1-\frac{2M}{r}+\frac{\eta \hat{\phi}_1^2}
{2 \Mpl^2 r^2}
+\frac{\eta M \hat{\phi}_1^2}
{2 \Mpl^2 r^3}
+\frac{2M \hat{\phi}_1 [\eta M \hat{\phi}_1
+24 \alpha \xi_{,\phi}(\phi_0)]}{3 \Mpl^2 r^4}
+{\cal O}(r^{-5})\,,
\label{hinf} \\
\phi &=& \phi_0+\frac{\hat{\phi}_1}{r}
+\frac{M\hat{\phi}_1}{r^2}
+\frac{(16 M^2 \Mpl^2-\eta \hat{\phi}_1^2)\hat{\phi}_1}
{12 \Mpl^2 r^3}
+\frac{M[6 \eta M^2 \Mpl^2 \hat{\phi}_1-\eta^2 \hat{\phi}_1^3
-12 \alpha M \Mpl^2 \xi_{,\phi}(\phi_0)]}
{3 \eta \Mpl^2 r^4}
+{\cal O}(r^{-5})\,,
\label{phiinf}
\ea
where we set $\hat{f}_1=-2M$.  
Then the quantities (\ref{cGdef})-(\ref{cFdef}) can be estimated as 
${\cal G}=\Mpl^2+8 \alpha M \xi_{,\phi}(\phi_0) \hat{\phi}_1 r^{-4}
+{\cal O}(r^{-5})$, 
${\cal H}=\Mpl^2+8 \alpha \xi_{,\phi}(\phi_0) \hat{\phi}_1 r^{-3}
+{\cal O}(r^{-4})$, and 
${\cal F}=\Mpl^2-16\alpha \xi_{,\phi}(\phi_0) \hat{\phi}_1 r^{-3}
+{\cal O}(r^{-4})$, so the squared radial and 
angular propagation speeds reduce, respectively, to 
\be
c_r^2=1+\frac{16 \alpha \xi_{,\phi}(\phi_0) \hat{\phi}_1}{\Mpl^2 r^3}
+{\cal O}(r^{-4})\,,\qquad 
c_\Omega^2=1-\frac{8 \alpha \xi_{,\phi}(\phi_0) \hat{\phi}_1}{\Mpl^2 r^3}
+{\cal O}(r^{-4})\,.\label{crOminf}
\ee
As $r \to \infty$, both $c_r^2$ and $c_{\Omega}^2$ 
approach 1. 

In the even-parity sector, the quantity ${\cal K}$ is 
expressed as 
\be
{\cal K}=2\eta \Mpl^4 \hat{\phi}_1^2+
\frac{4 \eta \Mpl^4 M \hat{\phi}_1^2}{r}
+{\cal O}(r^{-2})\,,
\label{Kexpan}
\ee
whose positivity is ensured for $\eta>0$. 
The radial propagation speed squared (\ref{cr3a}) yields 
\be
c_{r3}^2=1-\frac{128 \alpha^2 M \xi_{,\phi}^2(\phi_0) 
\hat{\phi}_1^2}{\Mpl^4 r^7}+{\cal O}(r^{-8})\,,
\label{cr3GB2}
\ee
which quickly approaches 1 at large distances 
even compared to $c_r^2$ and $c_{\Omega}^2$.
For the angular propagation, we obtain
\be
c_{\Omega \pm}^2=1 \pm \frac{4 \xi_{,\phi}(\phi_0) 
[(\eta \hat{\phi}_1^2+72 M^2 \Mpl^2)^{1/2} 
\mp \sqrt{\eta} \hat{\phi}_1]}
{\sqrt{\eta} \Mpl^2 r^3}|\alpha|+{\cal O}(r^{-4})\,.
\label{cOpm}
\ee
For $r \gg r_s$ all the squared propagation speeds given above 
rapidly approach 1, so the sign of $\alpha$ 
does not matter for the discussion of Laplacian instability. 
Provided that $\eta>0$, there are neither ghost nor 
Laplacian instabilities at spatial infinity.

\subsection{Numerical solutions and stability conditions}

The discussions in Secs.~\ref{censec} and \ref{infsec} 
show that, under the bound (\ref{alphacon}) 
with $\eta>0$, there are NS solutions with a nontrivial 
scalar profile consistent with all the linear stability 
conditions around $r=0$ and $r \to \infty$.  
However, they do not necessarily guarantee the linear stability 
of solutions at intermediate distances, so we will numerically 
study whether neither ghost nor Laplacian instabilities 
appear at any radius $r$.

For concreteness, we study the linear scalar-GB coupling given by the $n=1$ case of 
Eq.~\eqref{powerGB}. 
Since this corresponds to $G_2=\eta X$, $G_3=0$, $G_4=\Mpl^2/2$, 
and $G_5=-4\alpha \Mpl r_0^2 \ln |X|$, it belongs to 
a subclass of shift-symmetric Horndeski theories. 
The same scalar-GB coupling can also accommodate the 
dilatonic coupling $\xi(\phi) \propto {\rm e}^{\mu \phi}$ 
in the limit $|\mu \phi| \ll 1$. 
Since ${\cal P}_{\phi}=0$ in the scalar-field 
Eq.~(\ref{Ephi2}), we have $r^2 \sqrt{f/h}\,J^r=Q
={\rm constant}$ and hence 
\be
\frac{h}{f} \left[ \eta f r^2 \phi'
+4 \alpha \Mpl r_0^2 f' (h-1)  
\right]=Q\sqrt{\frac{f}{h}}\,.
\ee
To satisfy the boundary conditions of $f$, $h$, and 
$\phi'$ at $r=0$, we require that $Q=0$. 
Then, the field derivative can be expressed as 
\be
\phi'(r)=-\frac{4\alpha \Mpl r_0^2 f'(h-1)}
{\eta f r^2}\,.
\label{phidr}
\ee
Substituting Eqs.~(\ref{fr=0}) and (\ref{hr=0}) into 
Eq.~(\ref{phidr}) around $r=0$, we have $\phi'(r) \propto r$ 
as consistent with Eq.~(\ref{phir=0}).
At spatial infinity, using the expanded 
solutions (\ref{finf}) and (\ref{hinf}) in 
Eq.~(\ref{phidr}) leads to $\phi'(r) \propto r^{-5}$. 
Since the integration constant $Q$ corresponds to the scalar charge,
the choice of $Q=0$ means that $\hat{\phi}_1=0$ in the expansion
of Eq.~(\ref{phiinf}) and hence 
$\phi=\phi_0-4\alpha M^2 \xi_{,\phi}(\phi_0)
/(\eta r^4)$. In Eq.~(\ref{Kexpan}) the leading-order 
term of ${\cal K}$ vanishes, but it is replaced by 
${\cal K}=512 \alpha^2 M^4 \Mpl^4 \xi_{,\phi}
(\phi_0)^2/(\eta r^6)+{\cal O}(r^{-7})$. 
Then, the no-ghost condition is satisfied for $\eta>0$.

To perform the numerical integration, we introduce 
the following variables
\be
s=\ln \frac{r}{r_0}\,,\qquad 
{\cal M}(r)=4 \pi r \Mpl^2 (1-h)\,,
\qquad 
m(r)=\frac{3{\cal M}(r)}{4\pi r_0^3 \rho_0}\,,\qquad
y=\frac{\rho}{\rho_0}\,,\qquad 
z=\frac{P}{\rho_0}\,.
\ee
The ADM mass of star can be computed as 
\be
M \equiv {\cal M} (r \to \infty)=
2.5435 \times 10^2\,m_{\rm \infty}\,M_{\odot}\,,
\ee
where $m_{\infty} \equiv m(r \to \infty)$ and 
$M_{\odot}=1.9884 \times 10^{33}$~g is the solar mass.

For the NS EOS, we exploit the analytic representation 
of SLy EOS parametrized by 
\be
\xi=\log_{10} (\rho/{\rm g \cdot cm}^{-3})\,,\qquad 
\zeta=\log_{10} (P/{\rm dyn \cdot cm}^{-2})\,,
\label{xizeta}
\ee
where the relation between $\xi$ and $\zeta$ is  
given in Ref.~\cite{Haensel:2004nu}. 
In terms of $y$ and $z$, we can express $\xi$ and 
$\zeta$ as 
\be
\xi=\alpha_1+\alpha_2 \ln y\,,\qquad 
\zeta=\alpha_3+\alpha_2 \ln z\,,
\ee
where $\alpha_1=\ln( \rho_0/{\rm g\cdot cm^{-3}})/\ln10$, 
$\alpha_2=(\ln10)^{-1}$, 
and $\alpha_3=\ln( \rho_0 / {\rm dyn\cdot cm^{-2}})/\ln10$. 
Then, the EOS translates to 
\be
z=\exp \left[ \frac{\zeta(\xi)-\alpha_3}{\alpha_2} \right]\,,
\label{zy}
\ee
so that $z$ is known by the analytic representation of 
$\zeta(\xi)$. 
The continuity Eq.~(\ref{mattereq}) gives the 
differential equation for $y$, as
\be
\frac{{\rm d}y}{{\rm d}s}
=-\frac{y(y+z)}{2z} \left( \frac{{\rm d} \zeta}
{{\rm d} \xi} \right)^{-1}
\frac{1}{f} \frac{{\rm d}f}{{\rm d}s}\,.
\label{ys}
\ee
We derive the differential equations of $f$ and $h$
by solving Eqs.~(\ref{back1}), (\ref{back3}), and 
(\ref{Ephi2}) for $f''$, $h'$, and $\phi''$. 
We replace the first-order field derivative $\phi'$  
in the differential equations of $f$ and $h$
by using Eq.~(\ref{phidr}).

\begin{figure}[ht]
\begin{center}
\includegraphics[height=3.4in,width=3.5in]{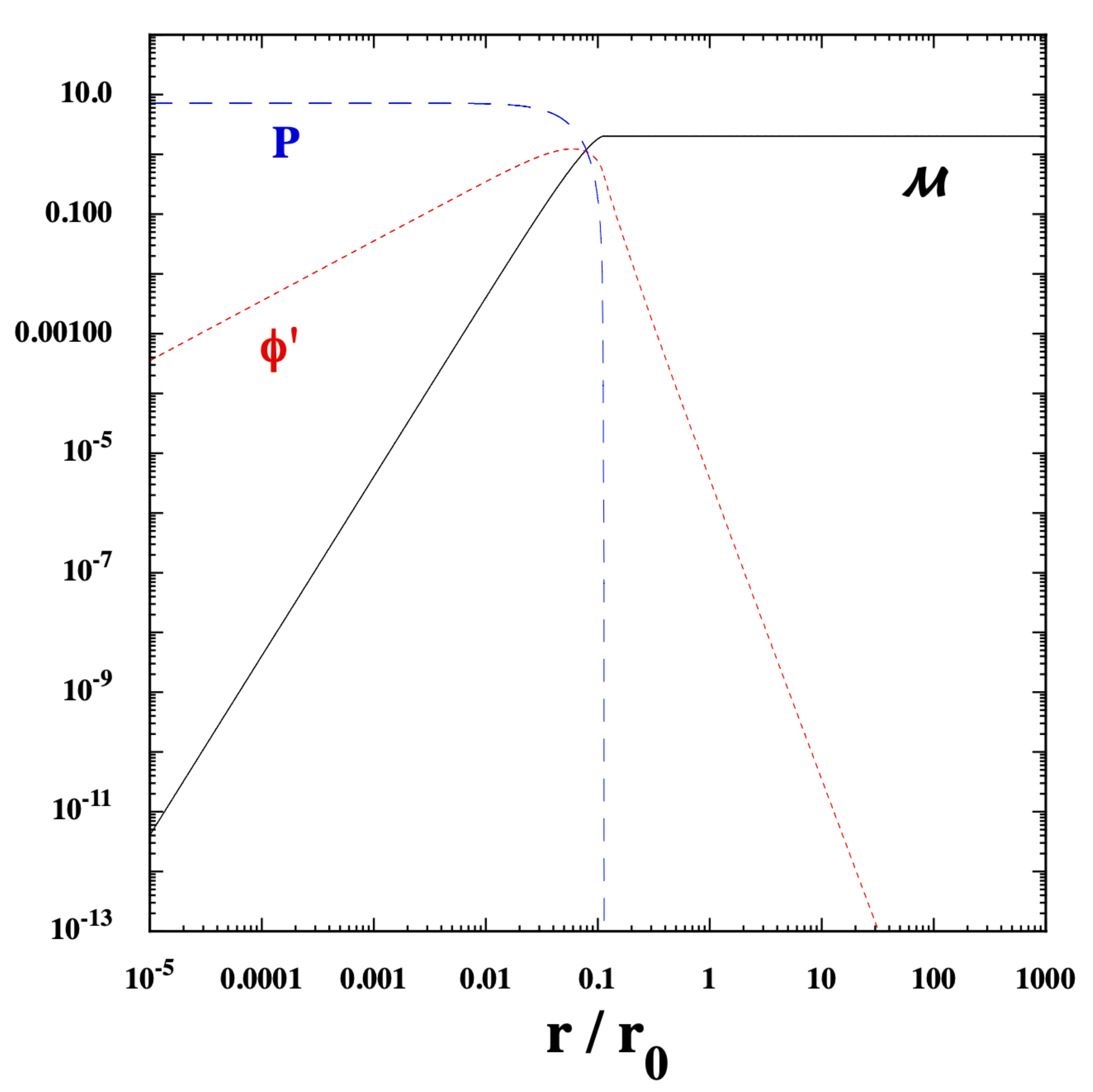}
\includegraphics[height=3.4in,width=3.5in]{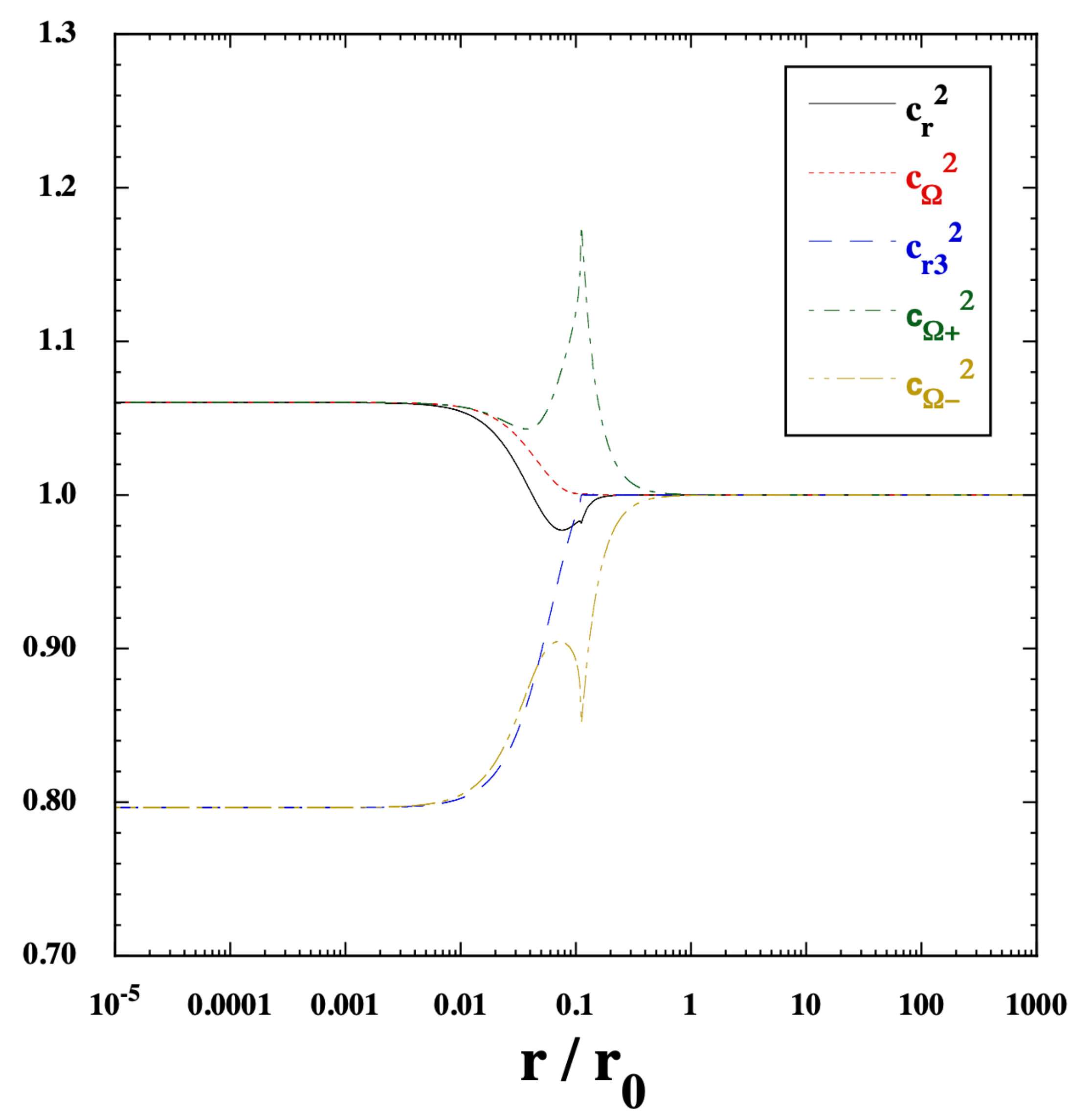}
\end{center}\vspace{-0.5cm}
\caption{\label{fig2}
(Left) Mass function ${\cal M}$ (normalized by the 
solar mass $M_{\odot}$), field derivative $\phi'$ 
(normalized by $\Mpl/r_0$), fluid pressure $P$ 
(normalized by $\rho_c$) versus $r/r_0$ for 
the linear scalar-GB coupling $\xi(\phi)=\Mpl r_0^2 \phi$
with $\alpha=2 \times 10^{-4}$ and $\eta=1$.
For the perfect fluid, we choose SLy EOS with 
the central density $\rho_c=15 \rho_0$. 
(Right) $c_r^2$, $c_\Omega^2$, $c_{r3}^2$, 
$c_{\Omega+}^2$, and $c_{\Omega -}^2$ versus $r/r_0$ 
for the same model parameters as those used 
in the left.
}
\end{figure}

In the left panel of Fig.~\ref{fig2}, we plot 
${\cal M}$, $\phi'$, and $P$ as a function of 
$r/r_0$ for $\alpha=2 \times 10^{-4}$, $\eta=1$, 
and $\rho_c=15 \rho_0$. We choose the boundary conditions 
at $r/r_0=10^{-5}$ to be consistent with Eqs.~(\ref{phir=0}) 
and (\ref{fr=0})-(\ref{Pr=0}).
As we estimated in Sec.~\ref{censec}, 
the field derivative increases as $\phi'(r) \propto r$ 
around $r=0$, with ${\cal M}(r) \propto r^3$ and $P(r)$  
decreasing according to Eq.~(\ref{Pr=0}). 
The radius at which $P$ vanishes is 
$r_s=0.114 r_0 \simeq 10.22$~km, which corresponds 
to the surface of star. The field derivative starts to 
decrease around $r=r_s$ and it has the dependence 
$\phi'(r) \propto r^{-5}$ at large distances. 
The growth of mass function ${\cal M}(r)$ saturates 
around the surface of star and it quickly approaches the ADM mass $M$.
In the numerical simulation of Fig.~\ref{fig2} we have $M=2.003 M_{\odot}$, 
which is smaller than the corresponding mass 
$2.044 M_{\odot}$ in GR ($\alpha=0$). 
This suppressed ADM mass in comparison to GR
is consistent with the result obtained for 
the dilatonic scalar-GB coupling 
$\xi(\phi) \propto {\rm e}^{\mu \phi}$ 
in Ref.~\cite{Pani:2011xm}.

In the right panel of Fig.~\ref{fig2}, we show $c_r^2$, 
$c_\Omega^2$, $c_{r3}^2$, $c_{\Omega+}^2$, and 
$c_{\Omega -}^2$ versus $r/r_0$ for the same model 
parameters as those used in the left.
As estimated from Eq.~(\ref{crOr=0}), the radial 
and angular propagation speeds in the odd-parity 
sector are superluminal in the regime $r \ll r_s$.  
In the even-parity sector the evolution of 
$c_{\Omega+}^2$ around $r=0$ is similar to 
$c_r^2$ and $c_{\Omega}^2$, whereas 
$c_{r3}^2$ and $c_{\Omega-}^2$ are in the subluminal range 
as estimated by Eq.~(\ref{cr3GB}).
Around $r=r_s$, $c_{\Omega+}^2$ and $c_{\Omega-}^2$ 
exhibit temporal increase and decrease, 
respectively, but they remain to be finite positive values
without Laplacian instabilities. 
Outside the star, all the squared 
propagation speeds shown in Fig.~\ref{fig2} 
quickly approach 1, as consistent with 
the discussion in Sec.~\ref{infsec}.
Numerically, we have also confirmed that the no-ghost 
conditions of odd- and even-parity perturbations 
are satisfied at any distance $r$.

The numerical simulation of Fig.~\ref{fig2} corresponds 
to $\rho_c=15 \rho_0$, $w_c=0.48$, $\eta=1$, and $n=1$, 
so the bound (\ref{alpocon}) gives 
$|\alpha|<3.19 \times 10^{-4}$. 
For the coupling $\alpha$ in this range, we numerically confirmed that 
all the linear stability conditions are satisfied at any distance $r$. 
Since $c_{r3}^2$ and $c_{\Omega -}^2$ are smallest at the center 
of NS, the upper limit of $|\alpha|$ is determined by their 
values at $r=0$ as we performed in Sec.~\ref{censec}.
When $\alpha<0$, the background solution and its linear stability 
are similar to those for the corresponding positive value $|\alpha|$. 
If $|\alpha|$ exceeds the upper limit $|\alpha_{\rm max}|$, there are 
Laplacian instabilities associated with negative values of 
$c_{r3}^2$ and $c_{\Omega -}^2$ at $r=0$. 
Thus, the theoretical bound (\ref{alpocon}) is sufficiently accurate for 
the estimation of maximum allowed values of $|\alpha|$. 
We have also performed numerical simulations 
for the scalar-GB couplings (\ref{powerGB}) with different 
powers $n$ and confirmed that, for $\alpha$ in the 
range (\ref{alpocon}) with $\eta>0$,  there are hairy 
NS solutions consistent with all the linear stability conditions.

\section{Scalar-Gauss-Bonnet and regular couplings}
\label{addsec}

In this section, we study the existence and the linear stability of 
NS solutions with a nontrivial scalar profile 
in the presence of several regular couplings 
besides the scalar-GB coupling 
$\alpha \xi(\phi) R_{\rm GB}^2$. 
We also take into account the Einstein-Hilbert term $\Mpl^2 R/2$ 
and the canonical kinetic term $\eta X$ (with $\eta>0$) in the action. 
We exploit SLy EOS for the numerical analysis 
in this section.

\subsection{Cubic Galileon and scalar-GB couplings}

The cubic Galileon corresponds to the coupling function 
$G_3 \supset \mu_3 X$, where $\mu_3$ is a constant.
Let us consider theories given by the action 
\be
{\cal S}=\int {\rm d}^4 x \sqrt{-g} 
\left[ \frac{\Mpl^2}{2}R+\eta X
+\alpha \xi(\phi)R_{\rm GB}^2
+\mu_3 X \square \phi \right]\,.
\label{actioncom1}
\ee

We first derive the solutions expanded around 
$r=0$ and then discuss the linear stability of them. 
In doing so, we use the approximation $|\alpha| \ll 1$ and pick up terms up to 
the order of $\alpha^2$.
Then, the coefficient $\phi_2$ in the field 
expansion (\ref{phir=0}) is 
\be
\phi_2=\frac{3 \eta \Mpl^6-32\rho_c^2 (1+2w_c)\xi_{,\phi}^2(\phi_c)\alpha^2}
{24 \mu_3 \Mpl^6} 
\left[ 1-\sqrt{1-\frac{32\Mpl^8 \rho_c^2 
(1+3w_c)\xi_{,\phi}(\phi_c) \mu_3\alpha}
{[3 \eta \Mpl^6-32\rho_c^2 (1+2w_c)\xi_{,\phi}^2(\phi_c)\alpha^2]^2}} 
\right]+{\cal O}(\alpha^3)\,,
\label{phi2G3}
\ee
where we have chosen the branch recovering 
Eq.~(\ref{phi2}) in the limit $\mu_3 \to 0$. 
We do not necessarily assume that the 
cubic Galileon coupling is of the same 
order as the scalar-GB coupling. 
The consistency of the small $\alpha$ expansion requires that
\be
\xi_{,\phi}(\phi_c) |\mu_3 \alpha|
\ll \frac{\eta^2 \Mpl^4}
{\rho_c^2 (1+3w_c)}\,,
\label{xipbo}
\ee
where we also assumed $\alpha^2 \rho_c^2 \xi_{,\phi}^2(\phi_c) 
\ll \eta \Mpl^6$.
Up to the order of $r^2$, the metric 
components and fluid pressure are the same 
forms as Eqs.~(\ref{fr=0})-(\ref{Pr=0}) with 
$\phi'(r)=2\phi_2 r$.
In the limit that $\alpha \to 0$, we have 
$\phi_2=0$ even for $\mu_3 \neq 0$. 
This shows that the cubic-order coupling alone does 
not give rise to NS solutions with 
$\phi'(r) \neq 0$. 
This is consistent with the no-hair argument of Ref.~\cite{Lehebel:2017fag} for regular couplings in shift-symmetric 
Horndeski theories.
In other words, the scalar-GB coupling is needed for the realization of NSs with a nontrivial scalar profile.

On using the above background solutions around 
$r=0$, we find that ${\cal G}$, ${\cal H}$, 
and ${\cal F}$ are 
of the same form as Eq.~(\ref{GHF}), where 
$x=\alpha \xi_{,\phi}(\phi_c)\phi_2/\Mpl^2$ and $\phi_2$ 
satisfying Eq.~(\ref{phi2G3}). 
Then, the absence of ghost/Laplacian instabilities in the odd-parity 
sector requires that $x<1/16$.
The squared propagation speeds $c_r^2$, $c_{\Omega}^2$ and 
$c_{\Omega+}^2$ at $r=0$ are equivalent to 
$1/(1-16x)$.
The other two squared propagation speeds at $r=0$, 
which are relevant to the radial and angular 
propagations of $\delta \phi$, are modified to 
\be
c_{r3}^2(r=0)=c_{\Omega -}^2 (r=0)
=\frac{\eta-8 \mu_3 \phi_2}{\eta-12\mu_3 \phi_2}
-\frac{64\rho_c^2 \xi_{,\phi}^2 (\phi_c)
[\eta (1+w_c)-2\mu_3 \phi_2(5+6w_c)]}
{3\Mpl^6 (\eta-12\mu_3 \phi_2)^2} \alpha^2
+{\cal O}(\alpha^4)\,.
\label{cr3O}
\ee
Expanding Eq.~(\ref{phi2G3}) with respect to 
$\alpha$ and using the leading-order solution 
$\phi_2=2\rho_c^2 (1+3w_c)
\xi_{,\phi}(\phi_c)\alpha/(9 \eta \Mpl^4)$, 
the inequality (\ref{xipbo}) 
translates to $\mu_3 \phi_2 \ll 2\eta/9$.
In the limit that $|\mu_3 \alpha| \ll 1$, the 
first term on the right hand-side of 
Eq.~(\ref{cr3O}) approaches 1. 
On the other hand, even if $|\alpha| \ll 1$, 
the product $|\mu_3 \phi_2|$ can be as 
large as the order $0.1 \eta$ for a large coupling $|\mu_3|$. To avoid that the leading-order term
of Eq.~(\ref{cr3O}) becomes negative, we require the condition 
$\eta-12 \mu_3 \phi_2>0$. 
On using the solution (\ref{phi2G3}) 
under the approximation 
$\alpha^2 \rho_c^2 \xi_{,\phi}^2(\phi_c) 
\ll \eta \Mpl^6$, this condition translates to 
\be
\xi_{,\phi}(\phi_c) \mu_3 \alpha
<\frac{\eta^2 \Mpl^4}{4 \rho_c^2 (1+3w_c)}\,.
\label{xipbo2}
\ee
Around $r=0$, the leading-order contribution to (\ref{Kcon}) is 
proportional to $r^6$, such that 
\be
{\cal K}=\frac{8 \Mpl^2 [3\Mpl^2 (1-16x)^3 \phi_2^2  
(\eta-12\mu_3 \phi_2)+32 \rho_c^2 x^2]}
{3(1-16x)} r^6+{\cal O}(r^7)\,.
\ee
Here, we have not used the expansion with respect to $\alpha$. 
Provided that $x<1/16$ and $\eta-12\mu_3 \phi_2>0$, 
the leading-order term of ${\cal K}$ is positive. 
{}From the above discussion, the linear stability of NSs around $r=0$ is 
ensured under the condition (\ref{xipbo2}) with $x<1/16$.

At large distances, the solutions to $f$ 
and $h$ are the same forms as Eqs.~(\ref{finf}) and (\ref{hinf}) up to 
the order of $r^{-4}$.
The solution to the scalar field 
is modified to 
\be
\phi = \phi_0+\frac{\hat{\phi}_1}{r}
+\frac{M\hat{\phi}_1}{r^2}
+\frac{(16 M^2 \Mpl^2-\eta \hat{\phi}_1^2)\hat{\phi}_1 }{12 \Mpl^2 r^3}
+\frac{3\Mpl^2 
[4\eta M^3 \hat{\phi}_1
-8M^2\xi_{,\phi}(\phi_0)\alpha-\mu_3 
\hat{\phi}_1^2]-2\eta^2 M \hat{\phi}_1^3}
{6 \eta \Mpl^2 r^4}
+{\cal O}(r^{-5})\,,
\ee
in which the cubic Galileon coupling appears 
at the order of $r^{-4}$.
On using these large-distance solutions, it follows that 
$c_r^2$, $c_{\Omega}^2$, and ${\cal K}$ are the same 
as those given in Eqs.~(\ref{crOminf}) and (\ref{Kexpan}) 
up to the next-to-leading order.
On the other hand, the other squared propagation speeds are modified to 
\ba
c_{r3}^2 &=& 1+\frac{4 \hat{\phi}_1 \mu_3}
{\eta r^3}+{\cal O} (r^{-4})\,,\\
c_{\Omega \pm}^2 &=& 1\pm 
\frac{\sqrt{16 \alpha^2 \eta \xi_{,\phi}^2 (\phi_0) (72 M^2 \Mpl^2+\eta \hat{\phi}_1^2)
+\mu_3 \Mpl^2 \hat{\phi}_1^2
[ \mu_3 \Mpl^2-8 \alpha \eta  
\xi_{,\phi}
(\phi_0)] } \mp \hat{\phi}_1 
[ \mu_3 \Mpl^2+4 \alpha \eta \xi_{,\phi}
(\phi_0)]}{\eta \Mpl^2 r^3}
\nonumber \\
& &
+{\cal O} (r^{-4})\,,
\ea
both of which approach 1 in the limit 
$r \to \infty$. 

To study the linear stability of NS solutions with a nontrivial scalar profile discussed above, 
we perform the numerical integration for the linear scalar-GB 
coupling $\xi(\phi)=\Mpl r_0^2 \phi$ in the presence of 
cubic Galileons.
Since this belongs to a subclass of shift-symmetric Horndeski theories, 
we have $J^r=(Q/r^2)\sqrt{f/h}$ from Eq.~(\ref{Ephi2}).
The regularity at $r=0$ leads to $Q=0$ and hence 
\be
\phi'(r)=\frac{\eta f r}{\mu_3 h (rf'+4f)}
\left[ 1-\sqrt{1+
\frac{8 \tilde{\mu}_3 \alpha r_0^4
f'h(rf'+4f)(h-1)}{\eta^2 r^3  f^2}} 
\right]\,,
\label{phirb}
\ee
where $\tilde{\mu}_3$ is a dimensionless coupling 
defined by 
\be
\tilde{\mu}_3 \equiv \mu_3 \frac{\Mpl}{r_0^2}\,.
\ee
In the limit $\tilde{\mu}_3 \to 0$, the branch (\ref{phirb}) 
smoothly approaches (\ref{phidr}).
In the following, we consider the positive scalar-GB 
coupling ($\alpha>0$) and study the effect of 
cubic Galileons on the linear stability of NS.  

{}From Eq.~(\ref{phi2G3}), the positive Galileon coupling 
$\tilde{\mu}_3>0$ leads to the enhancement of $\phi'(r)$ 
around $r=0$ in comparison to the case $\tilde{\mu}_3=0$. 
On the other hand, Eq.~(\ref{xipbo2}) gives 
the following upper bound
\be
\tilde{\mu}_3<\tilde{\mu}_3^{{\rm max}} 
\equiv \frac{\eta^2}
{256 \pi^2 \alpha} 
\left( \frac{\rho_0}{\rho_c} \right)^2 \frac{1}{1+3w_c}\,.
\label{tmu3bo2}
\ee
Due to this limit, the enhancement of $\phi'(r)$ induced by 
the positive $\tilde{\mu}_3$ is not so significant.
With the model parameters $\alpha=2.0 \times 10^{-4}$, 
$\rho_c=15 \rho_0$, $w_c=0.48$, and $\eta=1$, for 
example, we have $\tilde{\mu}_3^{{\rm max}}=3.6 \times 10^{-3}$ 
from Eq.~(\ref{tmu3bo2}). In this case, we numerically find 
that the Galileon coupling needs to be in the range 
$\tilde{\mu}_3<3.0 \times 10^{-3}$ to satisfy all the 
linear stability conditions at any distance $r$.
Thus, the condition (\ref{tmu3bo2}) gives a good approximate criterion for the existence 
of NSs with a nontrivial scalar profile 
without instabilities.
For $\tilde{\mu}_3=2.5 \times 10^{-3}$, the ADM mass 
and radius of NS are $M=1.998 M_{\odot}$ and $r_s=10.22$~km, 
respectively. They are similar to the values $M=2.003 M_{\odot}$ 
and $r_s=10.22$~km derived for $\tilde{\mu}_3=0$. 
For increasing $\tilde{\mu}_3$, the squared propagation 
speeds $c_{r3}^2$ and $c_{\Omega -}^2$ tend to be larger than 
those for $\tilde{\mu}_3$ because of the approach of the term 
$\eta-12 \mu_3 \phi_2$ to $+0$ in Eq.~(\ref{cr3O}). 
We note that $c_r^2$, $c_{\Omega}^2$, and $c_{\Omega}^2$ are 
not strongly affected by a positive coupling $\tilde{\mu}_3$. 

\begin{figure}[ht]
\begin{center}
\includegraphics[height=3.4in,width=3.5in]{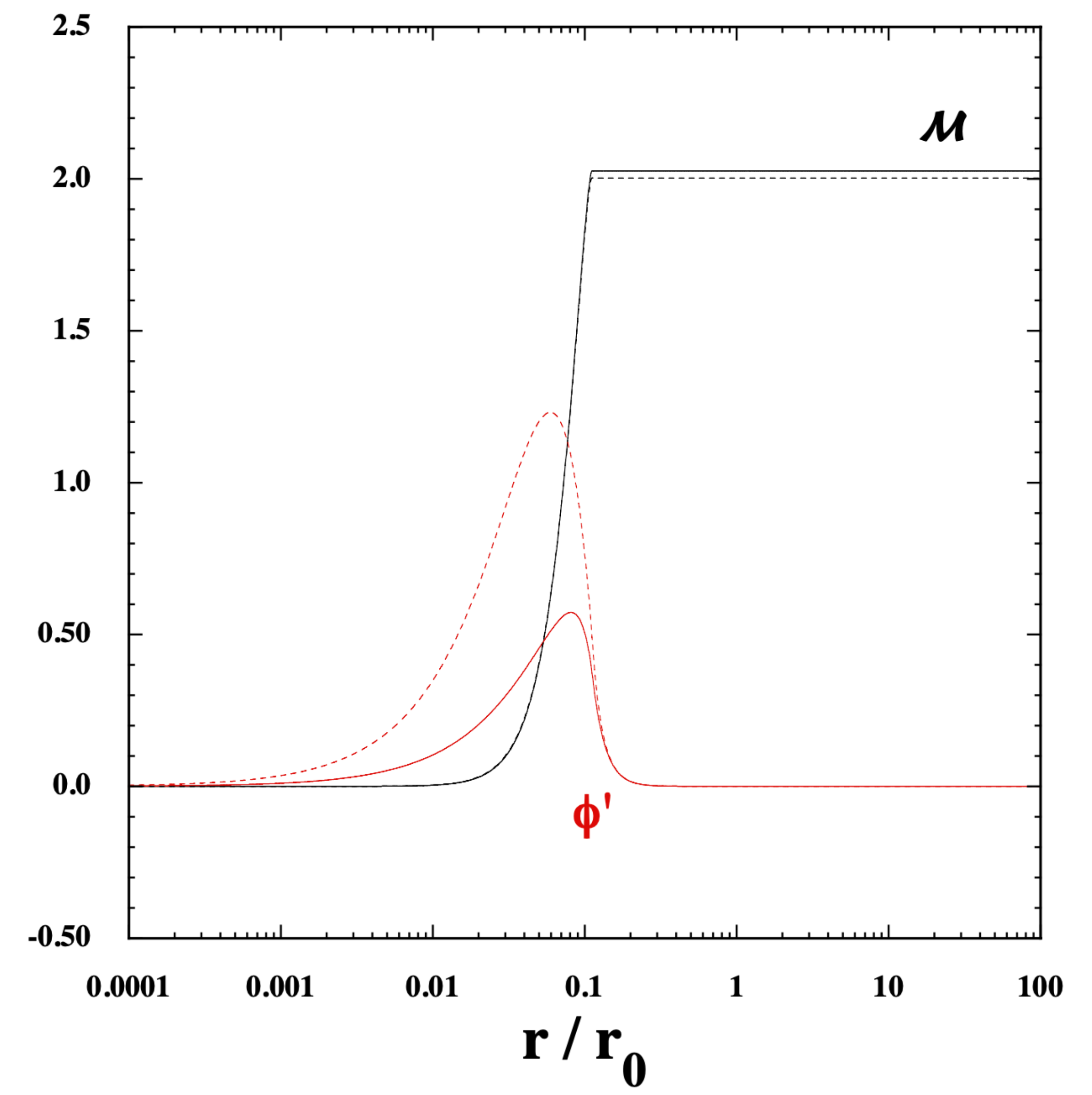}
\includegraphics[height=3.4in,width=3.5in]{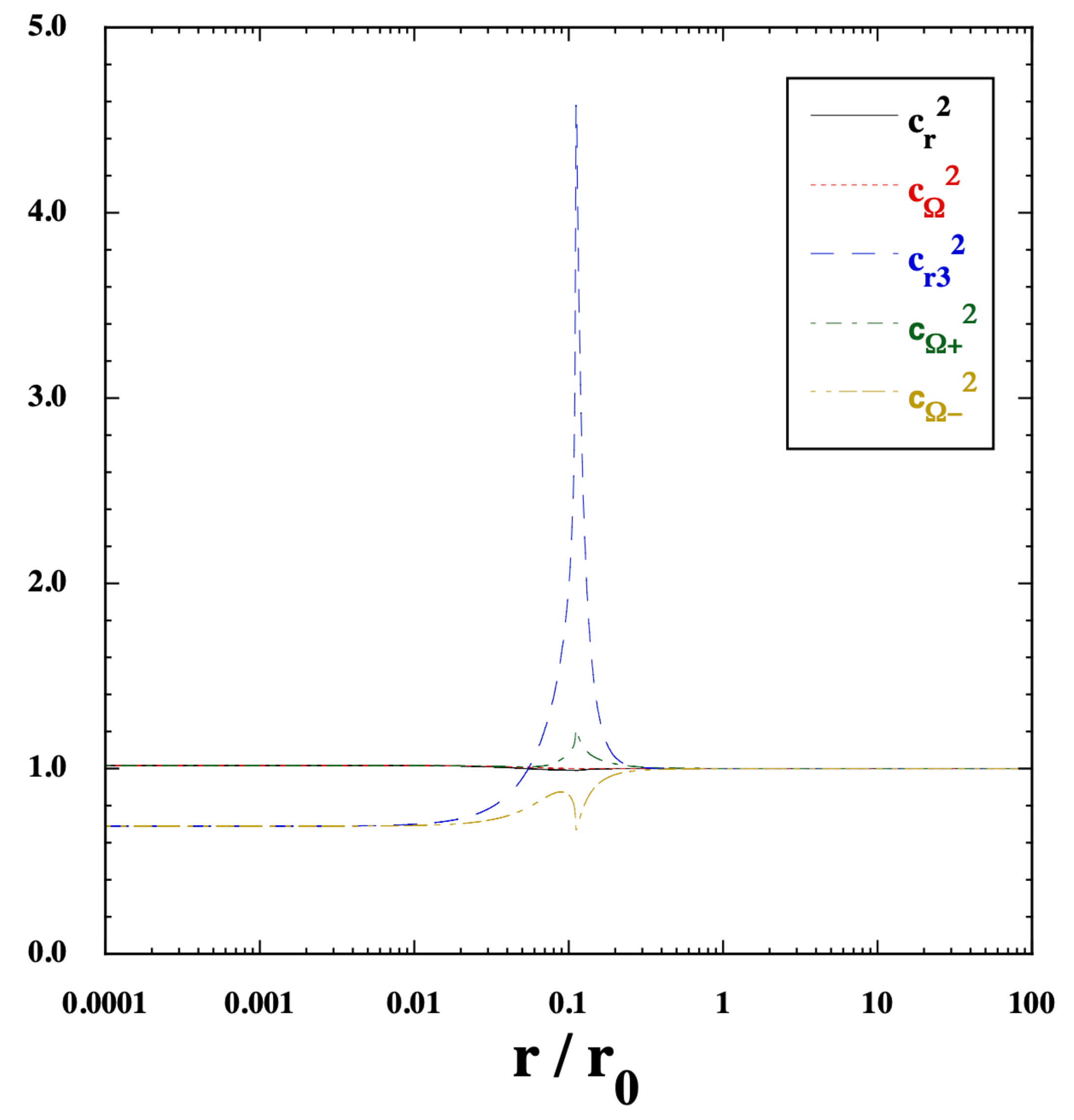}
\end{center}\vspace{-0.5cm}
\caption{\label{fig3}
(Left) The solid lines show the mass function 
${\cal M}$ (normalized by the solar mass $M_{\odot}$) 
and field derivative $\phi'$ (normalized by $\Mpl/r_0$) 
versus $r/r_0$ for the negative cubic Galileon coupling 
$\tilde{\mu}_3=-0.1$ with $\alpha=2 \times 10^{-4}$, 
$\eta=1$, and $\rho_c=15 \rho_0$. 
The dotted lines correspond to
${\cal M}$ and $\phi'$ for $\tilde{\mu}_3=0$, 
$\alpha=2 \times 10^{-4}$, $\eta=1$, and $\rho_c=15 \rho_0$, 
i.e., the case plotted in Fig.~\ref{fig2}. 
(Right) $c_r^2$, $c_\Omega^2$, $c_{r3}^2$, 
$c_{\Omega+}^2$, and $c_{\Omega -}^2$ versus $r/r_0$ 
for $\tilde{\mu}_3=-0.1$, $\alpha=2 \times 10^{-4}$, 
$\eta=1$, and $\rho_c=15 \rho_0$. 
}
\end{figure}

The negative value of $\tilde{\mu}_3$ is not bounded by Eq.~(\ref{tmu3bo2}). 
In the left panel of Fig.~\ref{fig3}, we plot ${\cal M}$ and $\phi'$ versus 
$r/r_0$ for $\tilde{\mu}_3=-0.1$ as solid lines. 
In comparison to the case $\tilde{\mu}_3=0$, the field 
derivative is suppressed by the negative Galileon coupling. 
For $\tilde{\mu}_3=-0.1$ the ADM mass is $M=2.026 M_{\odot}$, 
which is larger than $M=2.003 M_{\odot}$ derived for $\tilde{\mu}_3=0$. 
This increase of $M$ is attributed to the suppression of $\phi'$ 
induced by the negative $\tilde{\mu}_3$.
In the right panel of Fig.~\ref{fig3}, we show the five squared 
propagation speeds versus $r/r_0$ for $\tilde{\mu}_3=-0.1$. 
They are different from 1 inside the star, but all of them  
are larger than 0. In comparison to the case $\tilde{\mu}_3=0$ 
shown in Fig.~\ref{fig2}, there are more rapid temporal increase of $c_{r3}^2$ 
and decrease of $c_{\Omega-}^2$ around the surface of star. 
For decreasing $\tilde{\mu}_3$, we find that these transient 
variations of $c_{r3}^2$ and  $c_{\Omega-}^2$ tend to be 
more significant. 
Numerically, we find that there is a lower limit of $\tilde{\mu}_3$
in general to avoid that $c_{\Omega-}^2$ becomes negative 
around $r=r_s$.
With the model parameters $\alpha=2 \times 10^{-4}$, 
$\eta=1$, and $\rho_c=15 \rho_0$, for example, this instability 
arises for $\tilde{\mu}_3<\tilde{\mu}_3^{{\rm min}}=-0.22$. 

The above argument shows that, for the coupling $\tilde{\mu}_3$ 
between $\tilde{\mu}_3^{{\rm min}}$ and $\tilde{\mu}_3^{{\rm max}}$, 
there are NS solutions with a nontrivial scalar profile free from ghost/Laplacian 
instabilities. As $\tilde{\mu}_3$ approaches $\tilde{\mu}_3^{{\rm min}}$ 
or $\tilde{\mu}_3^{{\rm max}}$, the squared propagation speeds 
$c_{r3}^2$ and $c_{\Omega-}^2$ of $\delta \phi$
exhibit difference from those for $\tilde{\mu}_3=0$.

\subsection{Quartic derivative and GB couplings}

We proceed to a theory with the linear quartic derivative coupling 
$G_4 \supset \mu_4 X$ besides the scalar-GB 
coupling $\alpha \xi(\phi) R_{\rm GB}^2$.
We note that the quartic coupling $\mu_4 X$ in $G_4$
is equivalent to the quintic coupling $-\mu_4 \phi$ 
in $G_5$ \cite{KYY}.
This theory is given by the action 
\be
{\cal S}=\int {\rm d}^4 x \sqrt{-g} 
\left[ \frac{\Mpl^2}{2}R+\eta X
+\alpha \xi(\phi)R_{\rm GB}^2
+\mu_4 X R 
+\mu_4 \left\{ (\square \phi)^{2}
-(\nabla_{\mu}\nabla_{\nu} \phi)
(\nabla^{\mu}\nabla^{\nu} \phi) \right\}
 \right]\,.
\label{actioncom2}
\ee
The quartic derivative coupling alone does not give rise to asymptotically-flat 
NS solutions \cite{Lehebel:2017fag}, but the presence of scalar-GB couplings modifies 
this no-hair property.

Then, the solutions expanded around $r=0$ are given by 
\ba
f &=& f_c+\frac{f_c [ \Mpl^2 \rho_c (1+3w_c)
-48 \alpha \xi_{,\phi}(\phi_c) \phi_2 (w_c \rho_c-8 \mu_4 \phi_2^2) ]}
{6 [\Mpl^2-16 \alpha \xi_{,\phi}(\phi_c)\phi_2]^2}r^2
+{\cal O}(r^3)\,,\label{fr=0d} \\
h &=& 1-\frac{\rho_c+24 \mu_4 \phi_2^2}
{3[\Mpl^2-16 \alpha \xi_{,\phi}(\phi_c)\phi_2]}r^2+{\cal O}(r^3)\,,
\label{hr=0d}\\
P &=& P_c-\frac{\rho_c (1+w_c)[ \Mpl^2 \rho_c (1+3w_c)
-48 \alpha \xi_{,\phi}(\phi_c) \phi_2 ( w_c \rho_c -8 \mu_4 \phi_2^2) ]}
{12 [\Mpl^2-16 \alpha \xi_{,\phi}(\phi_c)\phi_2]^2}r^2
+{\cal O}(r^3)\,,
\label{Pr=0d}
\ea
where $\phi_2$ is the coefficient appearing in the 
scalar-field expansion $\phi=\phi_0+\phi_2 r^2+\cdots$.
Since the leading-order contribution to the 
scalar-field equation of motion
around $r=0$ is highly nonlinear in $\phi_2$,
it is impossible to solve it for any value $\alpha$.
So, we restrict ourselves to the case 
$|\alpha|\ll 1$.
For $|\alpha| \ll 1$, 
$\phi_2$ satisfies the cubic-order equation 
\be
\frac{9\Mpl^2 (\eta \Mpl^2+16 \mu_4^2 \phi_2^2-2w_c \rho_c \mu_4)}
{2\xi_{,\phi}(\phi_c)\rho_c (1+3w_c)(\rho_c+72 \mu_4 \phi_2^2)} \phi_2 
=\alpha\,,
\label{phi2thire}
\ee
at linear order in $\alpha$. 
In the limit $\alpha \to 0$ with a small derivative coupling constant $\mu_4$, the only solution to 
Eq.~(\ref{phi2thire}) is $\phi_2=0$.
Hence the nonvanishing scalar-GB coupling $\alpha$ is required to have the solution with a nontrivial scalar profile
with $\phi_2 \neq 0$.
On using these expanded solutions, the quantities associated with the stability of odd-parity perturbations at $r=0$ are of the same form as Eq.~(\ref{GHF}), where $x=\alpha \xi_{,\phi}(\phi_c)\phi_2/\Mpl^2$ and $\phi_2$ 
satisfies the relation (\ref{phi2thire}).
The stability in the odd-parity sector is ensured for $x<1/16$.

In the following, we will exploit the expansion with respect to the small coupling $\alpha$.
Around $r=0$, the no-ghost parameter of even-parity perturbations has the dependence
\be
{\cal K}= \left[ 8 \phi_2^2 \Mpl^2 \left( \eta \Mpl^2+144 \phi_2^2 \mu_4^2
+2 \rho_c \mu_4 \right)-256 \phi_2^3 \xi_{,\phi}(\phi_c) 
\left( \eta \Mpl^2+24 \phi_2^2 \mu_4^2
-\mu_4 \rho_c \right)\alpha+{\cal O}(\alpha^2) \right]r^6
+{\cal O}(r^7)\,.
\ee
To avoid ghosts for small $\alpha$, the leading-order term 
$8 \phi_2^2 \Mpl^2 ( \eta \Mpl^2+144 \phi_2^2 \mu_4^2
+2 \rho_c \mu_4 )r^6$ in ${\cal K}$
needs to be positive. This amounts to the condition 
\be
\mu_4 >-\frac{\eta \Mpl^2+144 \phi_2^2 \mu_4^2}{2\rho_c}\,,
\label{mu4no}
\ee
which indicates the existence of a lower bound on negative values of $\mu_4$. 
We note that $\phi_2$ depends on $\mu_4$ 
through Eq.~(\ref{phi2thire}). 
At $r=0$, $c_{\Omega+}^2$ as well as $c_r^2$ and $c_{\Omega}^2$ are of the same forms as Eq.~(\ref{crOr=0}), with $\phi_2$ satisfying Eq.~(\ref{phi2thire}).
The other two squared propagation speeds 
at $r=0$ are
\be
c_{r3}^2(r=0)=c_{\Omega -}^2 (r=0)
=\frac{\eta \Mpl^2+48 \phi_2^2 \mu_4^2
-2w_c \rho_c \mu_4}
{\eta \Mpl^2+144 \phi_2^2 \mu_4^2
+2 \rho_c \mu_4}+{\cal O}(\alpha)\,.
\label{cr3Od}
\ee
Under the condition (\ref{mu4no}), the absence of Laplacian instability requires that 
\be
\mu_4< \frac{\eta \Mpl^2+48 \phi_2^2 \mu_4^2}
{2w_c \rho_c}\,,
\label{mu4po}
\ee
which indicates the existence of 
an upper bound on positive values of $\mu_4$. 

The solutions to $f$ and $\phi$ expanded at large distances 
are of the same forms as Eqs.~(\ref{finf}) and 
(\ref{phiinf}), respectively, while the solution to $h$ is 
\be
h = 1-\frac{2M}{r}+\frac{\eta \hat{\phi}_1^2}
{2 \Mpl^2 r^2}
+\frac{\eta M \hat{\phi}_1^2}
{2 \Mpl^2 r^3}
+\frac{2\hat{\phi}_1 [ (M^2 \eta-3\mu_4) \hat{\phi}_1
+24 M\alpha \xi_{,\phi}(\phi_0)]}{3 \Mpl^2 r^4}
+{\cal O}(r^{-5})\,.
\label{hinfd}
\ee
Up to next-to-leading order, the quantities 
$c_r^2$, $c_{\Omega}^2$, ${\cal K}$, and $c_{\Omega \pm}^2$ are the same as those in Eqs.~(\ref{crOminf}), (\ref{Kexpan}), and 
(\ref{cOpm}), while the third radial propagation speed squared is modified to 
\be
c_{r3}^2=1-\frac{2\mu_4 \hat{\phi}_1^2}{\Mpl^2 r^4}
+{\cal O}(r^{-5})\,.
\ee
Provided that $\eta>0$, the linear stability conditions are consistently 
satisfied at spatial infinity.

For concreteness, we consider the linear scalar-GB coupling
$\xi(\phi)=\Mpl r_0^2 \phi$ besides the quartic derivative coupling. 
Since this theory falls in a subclass of shift-symmetric Horndeski theories, 
we have $J^r=(Q/r^2)\sqrt{f/h}$. The constant $Q$ must be 0 to satisfy 
the boundary conditions at $r=0$. Then, it follows that 
\be
\phi'(r)=-\frac{4\alpha f'(h-1)\Mpl r_0^2}
{\eta r^2 f-2\tilde{\mu}_4 r_0^2 
[rf'h+f(h-1)]}\,,
\label{phidr2}
\ee
where 
\be
\tilde{\mu}_4 \equiv \frac{\mu_4}{r_0^2}\,.
\ee
Substituting the expanded solutions (\ref{fr=0d}) and (\ref{hr=0d}) into Eq.~(\ref{phidr2}), we find 
that the field derivative behaves as 
$\phi'(r) \propto r$ around $r=0$.
Substitution of Eqs.~(\ref{finf}) and (\ref{hinfd}) into Eq.~(\ref{phidr2}) gives $\phi'(r)=16 \alpha M^2 \Mpl r_0^2/(\eta r^5)$ at spatial infinity
and hence $\hat{\phi}_1=0$ in Eq.~(\ref{phiinf}).
Note that the condition for the regularity at the center, $Q=0$, makes the leading scalar charge vanish at spatial infinity (${\hat \phi}_1=0$) in the expansion of Eq.~(\ref{phiinf}).

\begin{figure}[ht]
\begin{center}
\includegraphics[height=3.3in,width=3.4in]{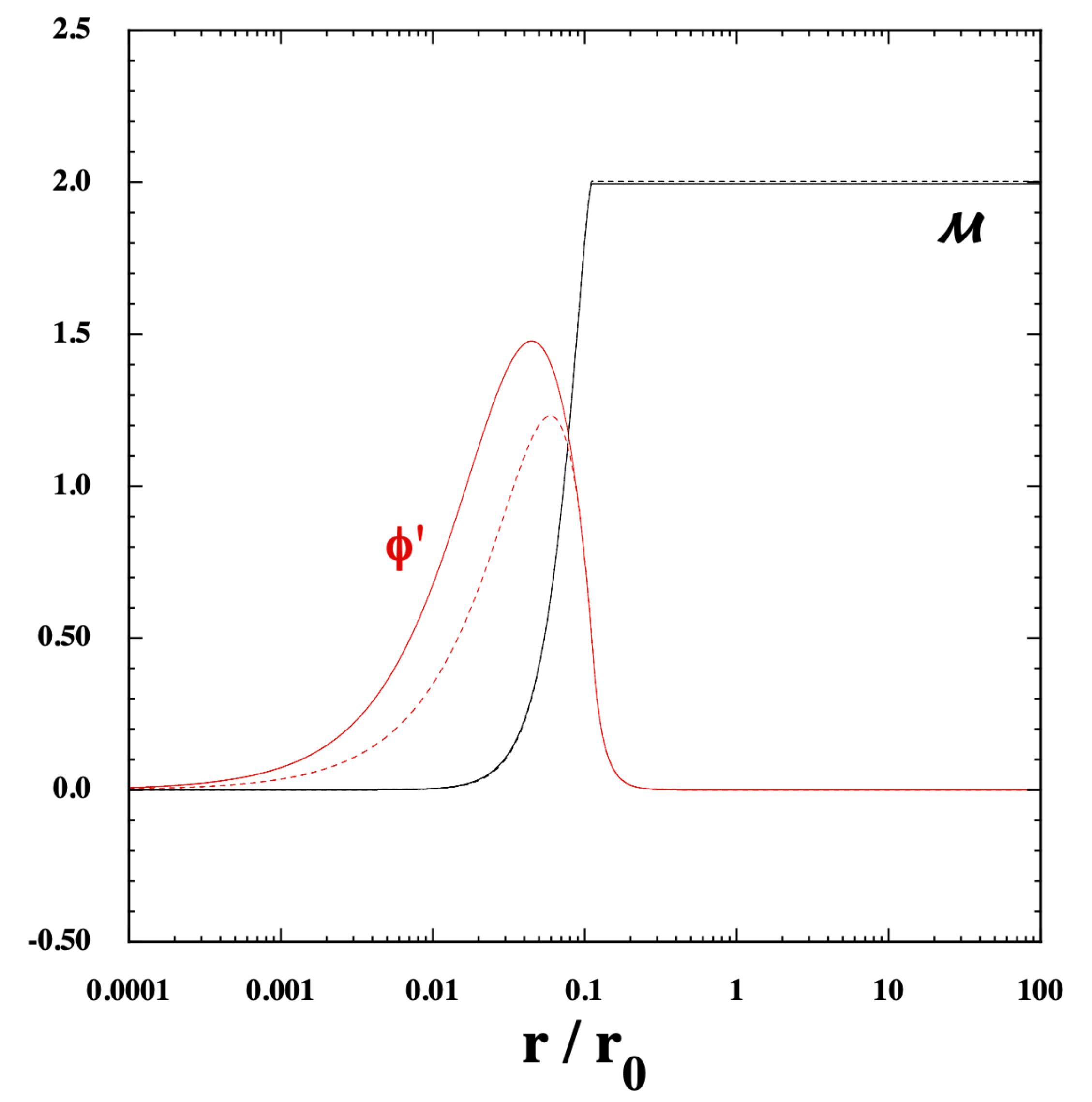}
\includegraphics[height=3.3in,width=3.3in]{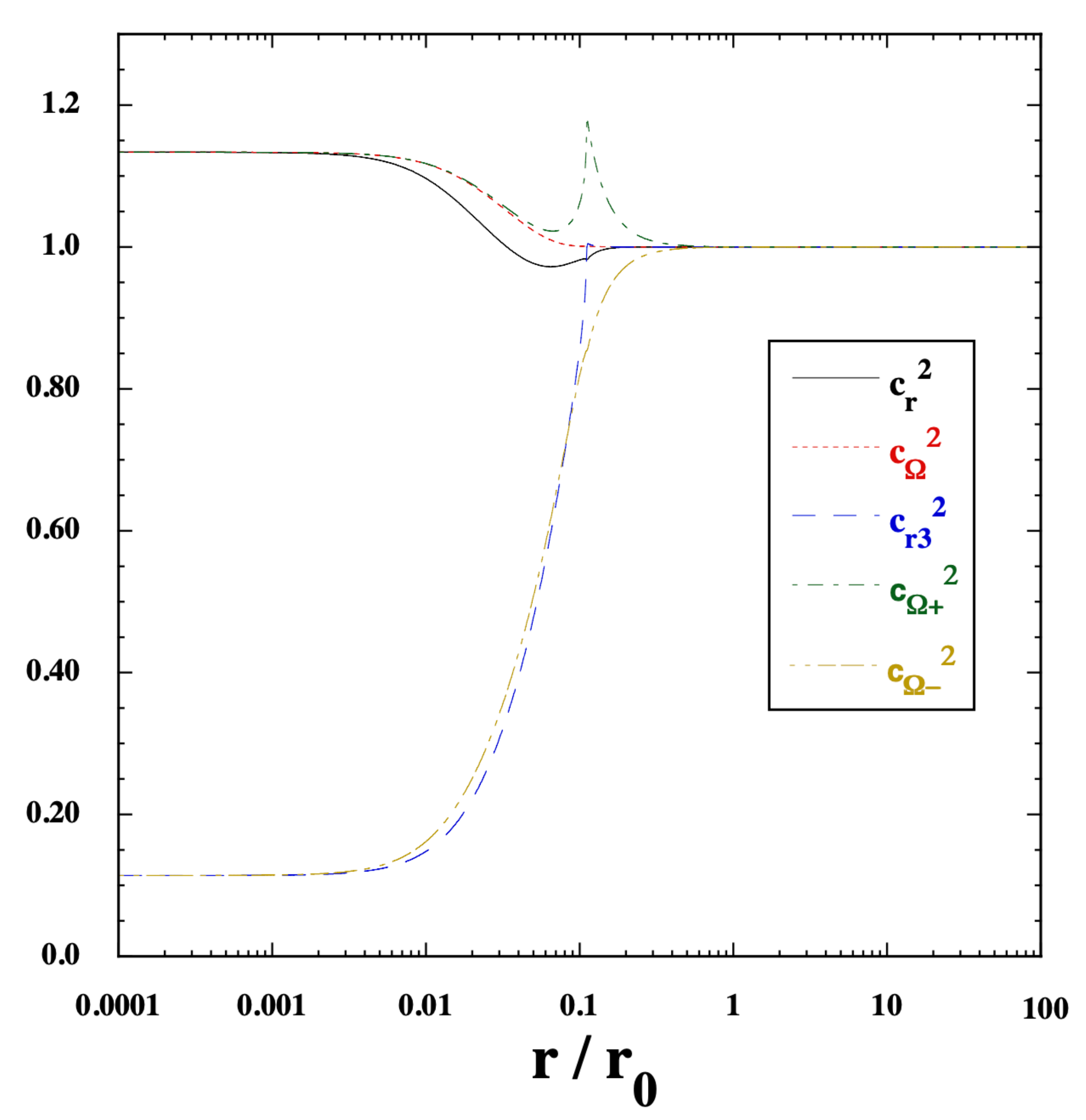}
\end{center}\vspace{-0.5cm}
\caption{\label{fig4}
(Left) Mass function ${\cal M}$ (normalized by the solar mass $M_{\odot}$) 
and field derivative $\phi'$ (normalized by $\Mpl/r_0$) 
versus $r/r_0$ for the quartic derivative coupling 
$\tilde{\mu}_4= 9 \times 10^{-4}$, 
$\alpha=2 \times 10^{-4}$, $\eta=1$, and $\rho_c=15 \rho_0$ (solid lines). 
The dotted lines show ${\cal M}$ and $\phi'$ for $\tilde{\mu}_4=0$, 
$\alpha=2 \times 10^{-4}$, $\eta=1$, and $\rho_c=15 \rho_0$.
(Right) $c_r^2$, $c_\Omega^2$, $c_{r3}^2$, 
$c_{\Omega+}^2$, and $c_{\Omega -}^2$ versus $r/r_0$ 
for $\tilde{\mu}_4= 9 \times 10^{-4}$, 
$\alpha=2 \times 10^{-4}$, $\eta=1$, and $\rho_c=15 \rho_0$.
}
\end{figure}

For given values of $\alpha$, $\rho_c$, $w_c$, and $\eta$, 
$\phi_2$ is known from Eq.~(\ref{phi2thire}) as a function of $\tilde{\mu}_4$. 
Then, the minimum value of $\tilde{\mu}_4$ can be found 
by Eq.~(\ref{mu4no}). 
When $\alpha=2 \times 10^{-4}$, $\rho_c=15 \rho_0$,
$w_c=0.48$, and $\eta=1$, we have
$\tilde{\mu}_4>-1.36 \times 10^{-3}$. 
Numerically, we find that the values of $c_{r3}^2$ and $c_{\Omega-}^2$ 
around $r=0$ become negative for $\tilde{\mu}_4<-1.30 \times 10^{-4}$. 
Hence the condition (\ref{mu4no}) gives a good approximate criterion for the existence of NS solutions with a nontrivial scalar profile consistent with linear stability conditions.
At the background level the negative coupling 
$\tilde{\mu}_4$ leads to tiny suppression of $\phi'(r)$, so the ADM mass of NS is only slightly increased. 
For $\tilde{\mu}_4=-1 \times 10^{-3}$ and $\alpha=2 \times 10^{-4}$, 
we obtain $M=2.008 M_{\odot}$, which is close to the value 
$M=2.003 M_{\odot}$ derived 
for $\tilde{\mu}_4=0$ and $\alpha=2 \times 10^{-4}$.
As $\mu_4$ approaches the lower bound (\ref{mu4no}), the leading-order contributions to $c_{r3}^2(r=0)$ and $c_{\Omega-}^2(r=0)$ become highly superluminal.
Outside the star, we numerically confirm that all of the propagation speeds quickly approach 1.  

The positive quartic coupling $\mu_4$ is constrained to be in the range satisfying (\ref{mu4po}). 
With the model parameters $\alpha=2 \times 10^{-4}$, $\rho_c=15 \rho_0$, 
$w_c=0.48$, and $\eta=1$, the condition (\ref{mu4po}) is satisfied for any positive $\mu_4$. However, we need to caution that the next-to-leading order correction to Eq.~(\ref{cr3Od}) gives rise to a negative term of order $-0.1$. 
For $\tilde{\mu}_4>1 \times 10^{-3}$, we numerically 
find that 
$c_{r3}^2$ and $c_{\Omega-}^2$ become negative around the center 
of star. Hence there is actually an upper bound of $\tilde{\mu}_4$ 
to avoid the Laplacian instability of even-parity perturbations. 
The plots in Fig.~\ref{fig4} correspond to the coupling 
$\tilde{\mu}_4=9 \times 10^{-4}$, in which case 
$c_{r3}^2$ and $c_{\Omega-}^2$ are as close as 0.1 
in the central region of star. 
Unlike the case $\tilde{\mu}_4=0$ shown in Fig.~\ref{fig2}, $c_{\Omega-}^2$ 
grows smoothly as a function of $r$ toward the asymptotic value 1. 
The behavior of other squared propagation speeds
$c_r^2$, $c_{\Omega}^2$, and $c_{\Omega+}^2$
is not much different from those seen in Fig.~\ref{fig2}. 
As we observe in the left panel of Fig.~\ref{fig4}, the field derivative inside the star is slightly enhanced in comparison to the case $\tilde{\mu}_4=0$. 
The resulting ADM mass $M=1.995 M_{\odot}$ is a bit 
smaller than the value $M=2.003 M_{\odot}$ obtained 
for $\tilde{\mu}_4=0$.

In summary, there are NS solutions with a nontrivial scalar profile free from ghost/Laplacian 
instabilities in certain ranges of the coupling $\tilde{\mu}_4$. 
For $\tilde{\mu}_4$ close to its lower limit, $c_{r3}^2$ and $c_{\Omega-}^2$ are
highly superluminal deep inside the star.
For $\tilde{\mu}_4$ close to its upper limit, $c_{r3}^2$ and $c_{\Omega-}^2$ 
approach $+0$ around $r=0$.

\subsection{Nonminimal Ricci scalar and GB couplings}
\label{nonminisec}

At the end of this section, we study the existence and stability of NS 
solutions in theories with nonminimal Ricci scalar and scalar-GB couplings. 
We incorporate a linear nonminimal coupling of the form 
$\lambda_4 \phi R$ in $G_4$. 
This also accommodates the dilatonic 
coupling ${\rm e}^{\lambda_4 \phi}R$ in the regime 
$|\lambda_4 \phi| \ll 1$. 
The action in such theories is given by 
\be
{\cal S}=\int {\rm d}^4 x \sqrt{-g} 
\left[ \frac{\Mpl^2}{2}R+\lambda_4 \phi R+\eta X
+\alpha \xi(\phi)R_{\rm GB}^2 \right]\,.
\label{actioncom1d}
\ee
Provided that the scalar-GB coupling is suppressed on a weak gravitational background, the dimensionless coupling 
constant $\tilde{\lambda}_4=\lambda_4/\Mpl$ is constrained to 
be $\tilde{\lambda}_4 \leq 2.5 \times 10^{-3}$ from 
Solar System experiments \cite{Tsujikawa:2008uc,DeFelice:2010aj}. 
On the strong gravitational background, 
we will study the effect of nonminimal coupling on the existence and the linear stability of NSs.

In the absence of the scalar-GB coupling, the squared propagation speeds 
are \cite{Kase:2021mix}
\be
c_{r}^2=c_{\Omega}^2=c_{r3}^2=c_{\Omega +}^2
=c_{\Omega -}^2=1\,,\qquad {\rm for} \qquad
\alpha=0\,, 
\label{cr2alp0}
\ee
at any distance $r$. 
These values are the same as those in GR, 
but the nonminimal coupling can give rise to 
NS solutions with a nontrivial scalar profile 
with $\phi'(r) \neq 0$ even in the absence of the scalar-GB coupling.

Let us consider the case in which 
both nonminimal and 
scalar-GB couplings are present.
The solutions expanded around $r=0$ are given by  
\ba
f &=& f_c+\frac{f_c \{ (\Mpl^2+2\lambda_4 \phi_c)
[\rho_c(1+3w_c)-12 \lambda_4 \phi_2]
-48 \alpha \xi_{,\phi}(\phi_c) \phi_2 (w_c \rho_c-8 \lambda_4 \phi_2)\}}
{6 [\Mpl^2+2\lambda_4 \phi_c-16 \alpha \xi_{,\phi}(\phi_c)\phi_2]^2}r^2
+{\cal O}(r^3)\,,\label{fr=0d2} \\
h &=& 1-\frac{\rho_c+12 \lambda_4 \phi_2}
{3[\Mpl^2+2\lambda_4 \phi_c-16 \alpha \xi_{,\phi}(\phi_c)\phi_2]}r^2+{\cal O}(r^3)\,,
\label{hr=0d2}\\
P &=& P_c-\frac{\rho_c(1+w_c)\{ (\Mpl^2+2\lambda_4 \phi_c)
[\rho_c(1+3w_c)-12 \lambda_4 \phi_2]
-48 \alpha \xi_{,\phi}(\phi_c) \phi_2 (w_c \rho_c-8 \lambda_4 \phi_2) \}}
{12 [\Mpl^2+2\lambda_4 \phi_c-16 \alpha \xi_{,\phi}(\phi_c)\phi_2]^2}r^2
+{\cal O}(r^3)\,,
\label{Pr=0d2}
\ea
where $\phi_2$ appears as a coefficient in Eq.~(\ref{phir=0}).
Performing the expansion with respect to a small scalar-GB coupling  
$\alpha$, $\phi_2$ obeys the second-order algebraic equation 
\be
\frac{3(\Mpl^2+2\lambda_4 \phi_c)\{ 6\eta \Mpl^2 \phi_2
+\lambda_4[\rho_c (1-3w_c)+12 \eta \phi_c \phi_2]
+36 \lambda_4^2 \phi_2\} }{4\xi_{,\phi}(\phi_c)
[\rho_c^2 (1+3w_c)+72 \lambda_4 \phi_2 
(w_c \rho_c-6 \lambda_4 \phi_2)]}=\alpha\,,
\ee
up to linear order in $\alpha$.
Choosing the regular branch in the limit $\alpha \to 0$, we obtain 
\be
\phi_2 (\alpha=0)=
\frac{\lambda_4\rho_c (3w_c-1)}{6[\eta \left( \Mpl^2
+2\lambda_4 \phi_c \right)+6\lambda_4^2]}\,.
\label{phi2al0}
\ee
Thus the nonminimal coupling $\lambda_4\neq 0$ 
alone leads to a nonvanishing value $\phi_2 (\alpha=0)\neq 0$ around the center of body.

At $r=0$, the absence of ghost/Laplacian instabilities for odd-parity perturbations requires that 
\ba
& &
{\cal G}(r=0)= 
\Mpl^2+2\lambda_4 \phi_c>0\,,
\label{nogon1}\\
& &
{\cal H}(r=0)={\cal F}(r=0)=
\Mpl^2+2\lambda_4 \phi_c-16 \alpha \xi_{,\phi}(\phi_c)\phi_2
>0\,, 
\label{noLap1}
\ea
with the associated squared propagation speeds 
\be
c_{r}^2(r=0)=c_{\Omega}^2(r=0)
=\left[ 1-\frac{16 \alpha \xi_{,\phi}(\phi_c)\phi_2}
{\Mpl^2+2\lambda_4 \phi_c} \right]^{-1}\,.
\label{cr2non}
\ee
Around $r=0$, the no-ghost parameter ${\cal K}$ is in proportion to $r^6$ and hence 
\be
{\cal K}/r^6 =
8 \phi_2^2 \left( \Mpl^2+2\lambda_4 \phi_c \right)
\left[ \eta \left( \Mpl^2+2\lambda_4 \phi_c \right)
+6\lambda_4^2 \right]
+{\cal O}(\alpha)\,,
\ee
where we performed the expansion 
with respect to small $\alpha$.
Under the condition (\ref{nogon1}), 
the absence of ghosts in the limit $\alpha \to 0$ requires that 
\be
\eta \left( \Mpl^2+2\lambda_4 \phi_c \right)
+6\lambda_4^2>0\,.
\label{nogon2}
\ee
Provided that $\eta>0$ and $\lambda_4 \phi_c>0$,
the conditions (\ref{nogon1}), (\ref{noLap1}), and (\ref{nogon2}) are automatically satisfied for $\alpha \to 0$. At $r=0$, we also obtain
\ba
c_{r3}^2(r=0) &=& c_{\Omega-}^2(r=0)= 
1-\frac{16 \lambda_4 \xi_{,\phi}(\phi_c) 
(\rho_c+P_c+6 \lambda_4 \phi_2)}
{(\Mpl^2+2\lambda_4 \phi_c)[\eta \left( \Mpl^2+2\lambda_4 \phi_c \right)
+6\lambda_4^2]}\alpha+{\cal O} (\alpha^2)\,,
\label{cr3non} \\
c_{\Omega+}^2(r=0) &=&1+\frac{16 \alpha \xi_{,\phi}(\phi_c)\phi_2}
{\Mpl^2+2\lambda_4 \phi_c}+{\cal O}(\alpha^2)\,.
\label{crOpnon} 
\ea
From Eqs.~(\ref{cr2non}) and (\ref{crOpnon}) 
we have $c_{r}^2(r=0)=c_{\Omega}^2(r=0)=c_{\Omega+}^2(r=0)$ up to linear order in $\alpha$. 
In the limit $\alpha \to 0$, all the above squared propagation speeds approach 1.

The solutions expanded at spatial infinity are given by
\ba
f &=& 1-\frac{2M}{r}+\frac{\lambda_4 \hat{\phi}_1
(4 \eta M \Mpl^2+8 \eta M \lambda_4\hat{\phi}_0 
+\eta \lambda_4\hat{\phi}_1+24 M \lambda_4^2)}
{(\Mpl^2+2\lambda_4 \hat{\phi}_0)
[\eta ( \Mpl^2+2\lambda_4 \hat{\phi}_0 )+6\lambda_4^2]r^2}
+{\cal O}(r^{-3})\,,\label{fla}\\
h &=& 1-\frac{2(M \Mpl^2+2M \lambda_4 \hat{\phi}_0 
-2\lambda_4 \hat{\phi}_1)}{(\Mpl^2+2\lambda_4 \phi_0)r}
+\frac{{\hat \phi}_1
[\eta^2 \Mpl^2 \hat{\phi}_1+2\eta \lambda_4
(\eta \hat{\phi}_0 \hat{\phi}_1+2M \Mpl^2)+2\eta \lambda_4^2 
(4M \hat{\phi}_0+\hat{\phi}_1)+24 M \lambda_4^3]}
{2(\Mpl^2+2\lambda_4 \hat{\phi}_0)
[\eta ( \Mpl^2+2\lambda_4 \hat{\phi}_0 )+6\lambda_4^2]r^2}
\nonumber \\
& &+{\cal O}(r^{-3})
\,,\label{hla}\\
\phi &=& 
\hat{\phi}_0+\frac{\hat{\phi}_1}{r}
+\frac{\hat{\phi}_1[2\eta M \Mpl^4+\eta \lambda_4 \Mpl^2 (8 M \hat{\phi}_0
-3\hat{\phi}_1)+\lambda_4^2 
\{ 2\eta  \hat{\phi}_0 (4M \hat{\phi}_0-3\hat{\phi}_1) 
+12M \Mpl^2 \}+12\lambda_4^3(2M \hat{\phi}_0-\hat{\phi}_1)]}
{2(\Mpl^2+2\lambda_4 \hat{\phi}_0)
[\eta ( \Mpl^2+2\lambda_4 \hat{\phi}_0 )+6\lambda_4^2]r^2}
\nonumber \\
& &
+{\cal O}(r^{-3})\,,
\label{phila}
\ea
where $M$, $\hat{\phi}_0$, $\hat{\phi}_1$ are integration constants. 
The coupling $\alpha$ does not appear in Eqs.~(\ref{fla})-(\ref{phila})
up to the order of $r^{-2}$, so the nonminimal coupling $\lambda_4$ provides larger contributions to $f$, $h$, and $\phi$ far outside 
the star in comparison to the scalar-GB term.
On using these large-distance solutions, the dominant contributions 
to ${\cal G}$, ${\cal H}$, and ${\cal F}$ are $\Mpl^2+2\lambda_4 \hat{\phi}_0$, 
so the linear stability conditions of odd-parity perturbations are satisfied 
if $\Mpl^2+2\lambda_4 \hat{\phi}_0>0$.
As for $c_r^2$ and $c_{\Omega}^2$, the term $\Mpl^2 r^3$ 
appearing in the denominators of Eq.~(\ref{crOminf}) is
modified to $(\Mpl^2+2\lambda_4 \hat{\phi}_0)r^3$.
In the even-parity sector, we have
\ba
{\cal K} &=& 2\hat{\phi}_1^2 \left( \Mpl^2+2\lambda_4 \hat{\phi}_0 
\right) \left[ \eta ( \Mpl^2+2\lambda_4 \hat{\phi}_0 )+6\lambda_4^2\right]+{\cal O} (r^{-1})\,,\\
c_{r3}^2 &=& 1-\frac{96 \alpha \xi_{,\phi}(\phi_c)\lambda_4^2\hat{\phi}_1}
{(\Mpl^2+2\lambda_4 \hat{\phi}_0 )[ \eta ( \Mpl^2+2\lambda_4 \hat{\phi}_0 )+6\lambda_4^2]r^3}+{\cal O}(r^{-4})\,.
\ea
Under the condition $\Mpl^2+2\lambda_4 \hat{\phi}_0>0$, 
the ghosts are absent for 
$\eta ( \Mpl^2+2\lambda_4 \hat{\phi}_0)+6\lambda_4^2>0$. 
In comparison to Eq.~(\ref{cr3GB2}), the nonminimal coupling gives 
rise to a term proportional to $r^{-3}$ in $c_{r3}^2-1$.
The expressions of squared angular propagation speeds 
$c_{\Omega \pm}^2$ are complicated, but they have the dependence 
$|c_{\Omega \pm}^2-1| \propto |\alpha|/r^3$ 
under the small $\alpha$ expansion.

\begin{figure}[ht]
\begin{center}
\includegraphics[height=3.3in,width=3.4in]{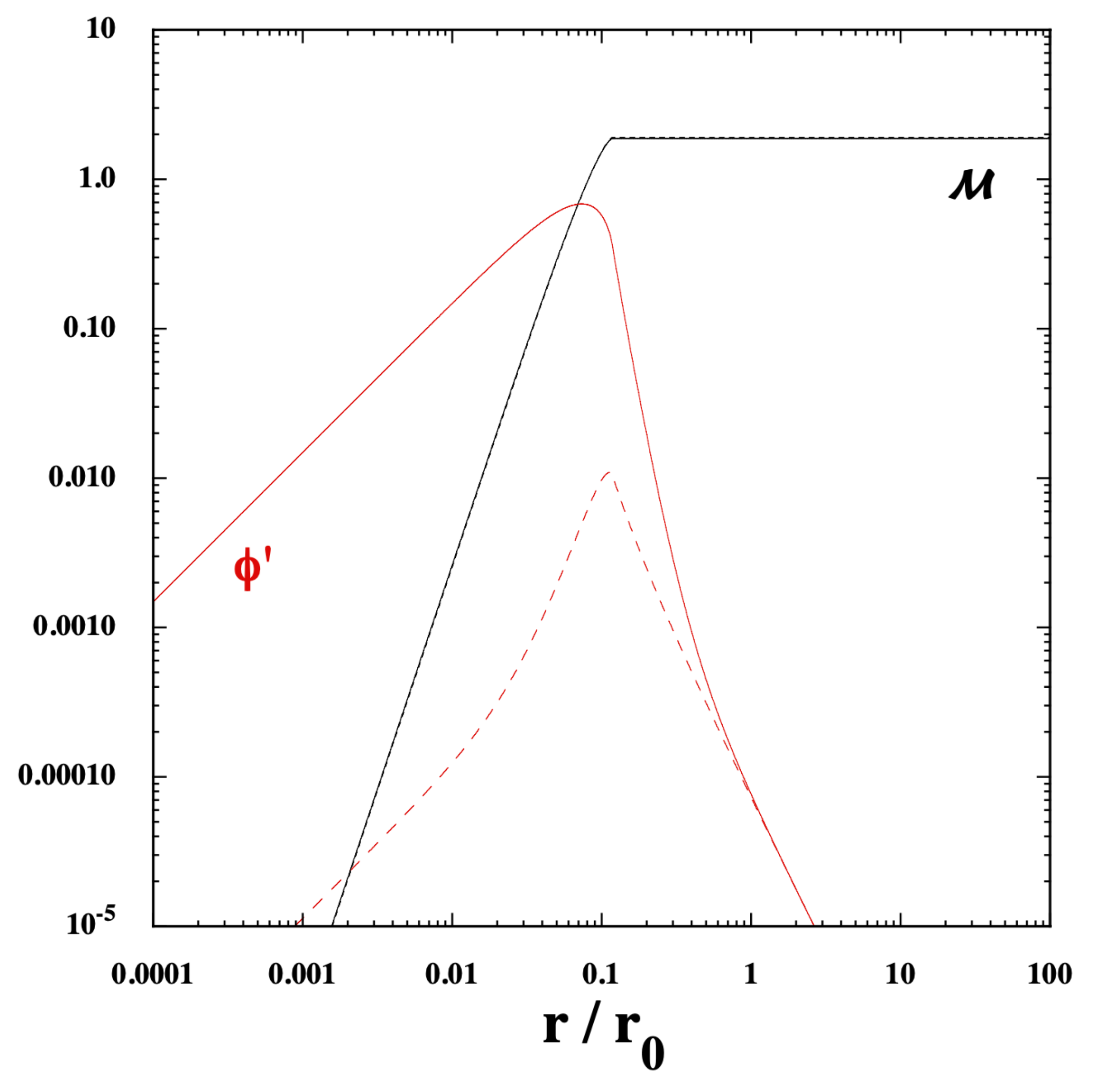}
\includegraphics[height=3.3in,width=3.3in]{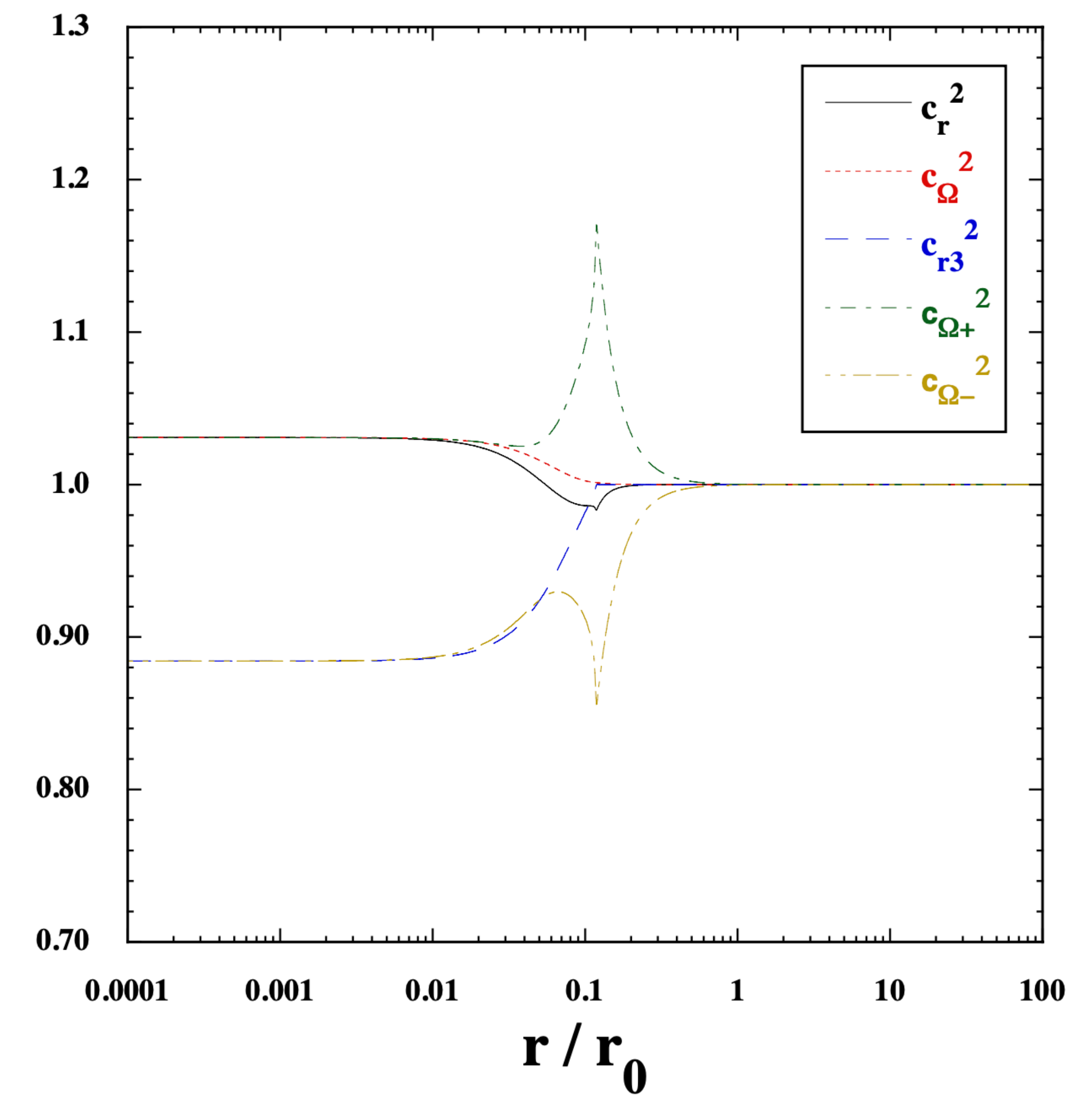}
\end{center}\vspace{-0.5cm}
\caption{\label{fig5}
(Left) Mass function ${\cal M}$ 
(normalized by the solar mass $M_{\odot}$) 
and field derivative $\phi'$ (normalized by $\Mpl/r_0$) 
versus $r/r_0$ for $\tilde{\lambda}_4=-2.5 \times 10^{-3}$, 
$\phi_c=\Mpl$, $\alpha=2.5 \times 10^{-4}$, $\eta=1$, 
and $\rho_c=10 \rho_0$ (solid lines).
The dashed lines correspond to the case $\alpha=0$, 
while the other model parameters are unchanged.
(Right) $c_r^2$, $c_\Omega^2$, $c_{r3}^2$, 
$c_{\Omega+}^2$, and $c_{\Omega -}^2$ versus 
$r/r_0$ for the same model parameters as those used 
for plotting solid lines in the left panel 
(i.e., both nonminimal and scalar-GB couplings are present). }
\end{figure}

In the left panel of Fig.~\ref{fig5}, we plot $\phi'$ versus $r/r_0$
for $\alpha=0$, $\tilde{\lambda}_4=\lambda_4/\Mpl=-2.5 \times 10^{-3}$, 
$\phi_c=\Mpl$, $\eta=1$, and $\rho_c=10 \rho_0$ as a dashed line.  
The field derivative increases around $r=0$ according to 
the relation $\phi' \simeq 2 \phi_2 r$, 
where $\phi_2$ is given by Eq.~(\ref{phi2al0}).
In Fig.~\ref{fig5}, we observe that $\phi'$ starts to decrease around the surface 
of star and it joins the large-distance solution 
$\phi' \simeq -\hat{\phi}_1/r^2$ with $\hat{\phi}_1<0$. 
Note that breaking the shift symmetry 
leads to the leading scalar charge 
$\hat{\phi}_1\neq 0$ even if the regular 
boundary conditions are imposed at the center.
This hairy NS solution satisfies all the 
linear stability conditions, 
with the squared propagation speeds given by Eq.~(\ref{cr2alp0}).
In this case, the ADM mass of NS is $M=1.908M_{\odot}$ with 
the radius $r_s=10.97$~km. 
They are almost similar to the values $M=1.912 M_{\odot}$ 
and $r_s=11.00$~km in GR with the same central density 
$\rho_c=10 \rho_0$.
Increasing the value of $|\lambda_4 \phi_c|$ further, it is possible to realize NSs with a nontrivial scalar profile whose mass and radius 
exhibit notable difference from those in GR. 
If we take into account Solar System constraints, however, the coupling $\tilde{\lambda}_4$ 
should be less than the order $10^{-3}$. Hence we do not consider the case in 
which the product $|\tilde{\lambda}_4 \phi_c|$ exceeds the order of $10^{-3}$.

If the scalar-GB coupling is present besides the nonminimal coupling, 
it is possible to realize NS solutions with a nontrivial scalar profile as well.
The solid curves in the left panel of Fig.~\ref{fig5} correspond to 
the radial dependence of $\phi'$ and ${\cal M}$ 
for $\alpha=2.5 \times 10^{-4}$, $\tilde{\lambda}_4=-2.5 \times 10^{-3}$, 
and $\phi_c=\Mpl$. Inside the NS, the field derivative is significantly 
enhanced in comparison to the case $\alpha=0$. 
Around the surface of star $\phi'$ starts to decrease rapidly, 
but it enters the region with the radial dependence 
$\phi' \simeq -\hat{\phi}_1/r^2$ for $r \gtrsim r_0$.
As we already mentioned, this latter property is attributed to 
the fact that the contribution to $\phi'$ from the nonminimal 
coupling dominates over that from the scalar-GB 
coupling at large distances. 
For $\alpha=2.5 \times 10^{-4}$ the mass and radius of NS are found to be $M=1.873M_{\odot}$ and $r_s=10.93$~km, both of which are slightly 
smaller than those for $\alpha=0$ mentioned above. 
In the right panel of Fig.~\ref{fig5}, we show the five 
squared propagation speeds versus $r/r_0$ for $\alpha=2.5 \times 10^{-4}$, 
$\tilde{\lambda}_4=-2.5 \times 10^{-3}$, and $\phi_c=\Mpl$.
In comparison to the case $\alpha=0$ where all the propagation 
speeds are equivalent to 1, they are different from 1 deep inside the NS
and approach 1 outside the star. 
For the model parameters used in Fig.~\ref{fig5}, there are neither 
ghost nor Laplacian instabilities for 
NS solutions with a nontrivial scalar profile. 

Provided that $|\tilde{\lambda}_4 \phi_c| \lesssim 10^{-3}$, 
the scalar-GB coupling $|\alpha|$ larger than the order $10^{-5}$ 
gives the dominant contribution to $\phi'$ inside the star. 
Then, for $|\alpha| \gtrsim 10^{-5}$, the background NS 
solution and its linear stability are not much different from 
those for the scalar-GB coupling alone discussed in Sec.~\ref{GBsec}.

\section{Regularized 4D-Einstein-Gauss-Bonnet gravity}
\label{4DEGBsec}

In this section, we study the linear stability of NS solutions with a nontrivial scalar profile
in so-called ``4DEGB gravity'' arising from 
the reduction of higher-dimensional GB theory 
to 4 dimensions.
If we consider the GB term ${\cal R}_{\rm GB}^2$ 
in spacetime dimensions $D$ higher than 4,  
the field equations of motion following from the Lagrangian 
$L=\sqrt{-g}\,\hat{\alpha}_{\rm GB}{\cal R}_{\rm GB}^2$ vanish in 4 dimensions \cite{Lovelock:1971yv}.
However, rescaling the GB coupling constant as 
$\hat{\alpha}_{\rm GB} \to \alpha_{\rm GB}/(D-4)$ allows a possibility 
for extracting contributions of the higher-dimensional GB term \cite{Glavan:2019inb}. 
Under such a rescaling, in the limit $D \to 4$, 
it is possible to construct a regularized 4-dimensional theory by adding 
a counter-term to eliminate 
divergent parts of the theory \cite{Fernandes:2020nbq,Hennigar:2020lsl}. 
The other equivalent procedure is 
to perform a Kaluza-Klein reduction of $D$-dimensional Einstein-GB 
gravity on a $(D-4)$-dimensional 
maximally symmetric space with 
a vanishing spatial curvature \cite{Lu:2020iav,Kobayashi:2020wqy}. 
The size of such a maximally symmetric space
is characterized by a scalar field $\phi$.
The 4-dimensional action obtained from the Kaluza-Klein 
reduction of $D$-dimensional Einstein-GB theory belongs to 
a subclass of shift-symmetric Horndeski theories 
given by the coupling functions
\be
G_2=8 \alpha_{\rm GB} X^2\,,\qquad 
G_3=8 \alpha_{\rm GB} X\,,\qquad 
G_4=1+4\alpha_{\rm GB}X\,,\qquad 
G_5=4 \alpha_{\rm GB} \ln |X|\,.
\ee
Notice that the standard kinetic term $X$ is absent in $G_2$.
Throughout this section, we use the unit $\Mpl^2/2=1$.

{}From the scalar-field Eq.~(\ref{Ephi2}), we have 
$r^2 \sqrt{f/h}\,J^r=Q={\rm constant}$ and hence
\be
4\sqrt{\frac{h}{f}} \left( f'+2\phi' f \right) 
\left[ 1- h (1+r\phi')^2 \right]\alpha_{\rm GB}=Q\,.
\label{fieldeqGB}
\ee
Due to the regularity conditions $f'(0)=0$ and $\phi'(0)=0$ 
at the center of star, it follows that $Q=0$. 
To satisfy Eq.~(\ref{fieldeqGB}) with $Q=0$ at any radius $r$, 
we require that $1-h (1+r\phi')^2=0$.
The branch where $\phi(r)$ decreases at spatial infinity 
(satisfying the asymptotic flatness $h \to 1$ as $r \to \infty$) 
is given by 
\be
\phi'=\frac{1}{r} \left( \frac{1}{\sqrt{h}}-1 \right)\,.
\label{phiexp}
\ee
Taking the $r$ derivative of this equation and substituting 
$\phi''$ and Eq.~(\ref{phiexp}) into Eqs.~(\ref{back1}) and (\ref{back2}), 
we obtain
\ba
h' &=& -\frac{2(h-1)[ r^2+(h-1)\alpha_{\rm GB}]
+\rho r^4}{2r[r^2-2 (h-1) \alpha_{\rm GB}]}\,,\label{h4DGB}\\
f' &=& -f \frac{2(h-1)[r^2+(h-1)\alpha_{\rm GB}]
-P r^4}{2hr [r^2-2 (h-1) \alpha_{\rm GB}]}\,.\label{f4DGB}
\ea

Outside the star ($\rho=0=P$), there is the following 
analytic solution \cite{Lu:2020iav,Fernandes:2021ysi}
\be
h=f=1+\frac{r^2}{2\alpha_{\rm GB}} 
\left[ 1-\sqrt{1+\frac{8\alpha_{\rm GB} M}{r^3}} 
\right]\,,
\label{hfso}
\ee
where $M$ is an integration constant. 
At spatial infinity, this has the asymptotic behavior 
$h=f=1-2M/r+{\cal O}(r^{-4})$ and $\phi'=M/r^2+{\cal O}(r^{-3})$.

Inside the NS, the integrated solutions to $h$ and $f$ depend 
on the fluid EOS. For constant density $\rho$, we have 
the following exact solution \cite{Doneva:2020ped}
\ba
h &=& 1-\zeta r^2\,,\\
f &=& \frac{(1-\alpha_{\rm GB} \zeta)^2}{4(1+2\alpha_{\rm GB} \zeta)^2}
\left[ \frac{3(1+\alpha_{\rm GB} \zeta)}{1-\alpha_{\rm GB} \zeta} \sqrt{1-\zeta r_s^2} 
-\sqrt{1-\zeta r^2} \right]^2\,,\\
P &=& \rho \frac{(1-\alpha_{\rm GB} \zeta) 
[\sqrt{1-\zeta r^2}-\sqrt{1-\zeta r_s^2}]}{3(1+\alpha_{\rm GB} \zeta) \sqrt{1-\zeta r_s^2}
-(1-\alpha_{\rm GB} \zeta) \sqrt{1-\zeta r^2}}
=\rho \left(
\sqrt{\frac{1-\zeta r_s^2}{f}}
-1
\right)
\,,
\label{Pins}
\ea
where 
\be
\zeta \equiv 
\frac{1}{2\alpha_{\rm GB}} \left( \sqrt{1+\frac23 \rho \alpha_{\rm GB}} 
-1 \right)\,.
\ee
At the surface of star ($r=r_s$), the fluid pressure (\ref{Pins}) vanishes. 
Matching $h$ and $f$ with Eq.~(\ref{hfso}) at $r=r_s$, 
there is the relation $M=\rho r_s^3/12$. 
In the following, we will study the linear stability of NSs without assuming 
their EOSs.

The quantities associated with the linear stability against odd-parity 
perturbations are given by 
\ba
{\cal G} &=& \frac{2\sqrt{h}\,r^4-2r^2 (\sqrt{h}-1) 
(Pr^2+2-2h+4\sqrt{h}) \alpha_{\rm GB}-4 (\sqrt{h}+1)  (\sqrt{h}-1)^4 
\alpha_{\rm GB}^2}{r^2 \sqrt{h}\,[r^2+2(1-h)\alpha_{\rm GB}]}\,,\\
{\cal F} &=& \frac{2\{ r^4+r^2 [\rho r^2+2(h-1)] \alpha_{\rm GB}
-2(h-1)^2 \alpha_{\rm GB}^2 \} }{r^2[r^2+2(1-h)\alpha_{\rm GB}]}\,,\\
{\cal H} &=& 2+\frac{4(1-h)}{r^2} \alpha_{\rm GB}\,.
\ea
In the limit that $\alpha_{\rm GB} \to 0$ we have ${\cal G}={\cal F}={\cal H}=2$, 
so the linear stability against odd-parity perturbations is ensured for the small GB coupling.
On using the background Eqs.~(\ref{h4DGB}) and (\ref{f4DGB}), 
it follows that 
\be
{\cal K}=0\,,
\ee
at any radius $r>0$.
Inside the star, the term $(2{\cal P}_1-{\cal F})h \mu^2$ 
exactly cancels the contribution $-2{\cal H} ^2 r^4 (\rho+P)$, 
while, outside the star, ${\cal K}=(2{\cal P}_1-{\cal F})h \mu^2=0$. 
The fact that ${\cal K}$ vanishes everywhere, which is mostly 
attributed to the absence of a standard kinetic term in $G_2$, 
is the signal of a strong coupling problem. 
Indeed, the squared radial propagation speed
associated with the stability of scalar-field perturbation 
in the even-parity sector yields
\be
c_{r3}^2 \to \infty\,.
\ee
This divergent property of $c_{r3}^2$ arises from the existence 
of term ${\cal K}$ in the denominator of Eq.~(\ref{cr3}).
We note that the product ${\cal K}c_{r3}^2$ is finite. 
Exploiting the exact solution (\ref{hfso}) outside the star
and performing the expansion with respect to $1/r$ at 
spatial infinity, we obtain the dependence 
\be
{\cal K}c_{r3}^2=-\frac{256 M^3 \alpha_{\rm GB}}
{r^3}+{\cal O}(r^{-4})\,.
\label{Kcr3}
\ee
For $\alpha_{\rm GB}<0$, the leading-order contribution 
to ${\cal K}c_{r3}^2$ is positive at large distances. 

The quantities $B_1$ and $B_2$ also diverge, so 
this leads to the divergence of angular propagation speeds
in the even-parity sector. 
Expanding the product ${\cal K}B_2$ at spatial infinity, 
it follows that 
\be
{\cal K}B_2=\frac{128 M^3 \alpha_{\rm GB}}{r^3}
+{\cal O}(r^{-4})\,.
\label{KB2}
\ee
Dividing Eq.~(\ref{Kcr3}) by Eq.~(\ref{KB2}), we obtain 
\be
\frac{c_{r3}^2}{B_2}=-2+{\cal O}(r^{-1})\,.
\ee
Since the signs of $c_{r3}^2$ and $B_2$ are different from 
each other, either of the linear stability conditions 
(\ref{cr3}) or (\ref{B12con}) is violated at large distances.
Thus, the NS solutions in 4DEGB gravity not only suffer from 
the strong coupling problem but also the Laplacian 
instability of even-parity perturbations. 
These problems also persist for hairy BHs given by 
the line element (\ref{hfso}) present in 4DEGB theory. 
In the case of BHs, there is also the instability of even-parity 
perturbations in the vicinity of the event horizon \cite{Tsujikawa:2022lww}.

\section{$F(R_{\rm GB}^2)$ gravity}
\label{fGsec}

Finally, we study a modified GB gravity in which the 4-dimensional 
action contains an arbitrary function $F$ of the GB term 
$R_{\rm GB}^2$ besides the Einstein-Hilbert term.
This theory is given by the action 
\be
{\cal S}=\int {\rm d}^4 x \sqrt{-g} 
\left[ \frac{\Mpl^2}{2}R+F(R_{\rm GB}^2) \right]\,,
\label{LHG}
\ee
which is equivalent to \cite{KYY}
\be
{\cal S}=\int {\rm d}^4 x \sqrt{-g} 
\left[ \frac{\Mpl^2}{2}R+\xi(\phi) R_{\rm GB}^2 
-V(\phi) \right]\,,
\label{LHG2}
\ee
where 
\be
\varphi \equiv R_{\rm GB}^2\,,\qquad
\phi \equiv \Mpl r_0^4 \varphi\,,\qquad  
\xi(\phi) \equiv F_{,\varphi}\,,\qquad 
V(\phi) \equiv \varphi \xi-F\,.
\ee
The action (\ref{LHG2}) belongs to a subclass of Horndeski 
theories given by the coupling functions
\ba
& &
G_2=-V(\phi)+8 \xi^{(4)}(\phi) X^2 (3-\ln |X|)\,,\qquad G_3=4\xi^{(3)}(\phi) X (7-3\ln |X|)\,,\nonumber \\
& &
G_4=\frac{\Mpl^2}{2}+4\xi^{(2)}(\phi) X (2-\ln |X|)\,,\qquad
G_5=-4\xi^{(1)}(\phi) \ln |X|\,.
\ea
Hence there is no standard scalar kinetic term in $F(R_{\rm GB}^2)$ gravity. 

Let us focus on the power-law $F(R_{\rm GB}^2)$ models given by 
\be
F(R_{\rm GB}^2)=\beta \left( R_{\rm GB} \right)^n\,,
\label{powerlaw}
\ee
where $\beta$ and $n$ are constants. 
We consider the positive integers $n$ in the range $n \geq 2$.
Then, the GB coupling function $\xi(\phi)$ and scalar potential $V(\phi)$ 
in the action (\ref{LHG2}) yield
\be
\xi(\phi)=\alpha n \frac{r_0^2 \Mpl^2}{2}
\left( \frac{\phi}{\Mpl} \right)^{n-1}\,,\qquad
V(\phi)=\alpha (n-1) \frac{\Mpl^2}{2r_0^2} 
\left( \frac{\phi}{\Mpl} \right)^n\,,
\label{xiV}
\ee
where $\alpha \equiv 2r_0^{2-4n} \Mpl^{-2} \beta$ is 
a dimensionless coupling.

{}From the scalar-field Eq.~(\ref{Ephi2}), we obtain
\be
\alpha \phi^{n-2} \left\{ r^2 f^2 \phi 
-2 \left[ ff'h' (3h-1)+f'^2 h (1-h)+2f'' f h (h-1) \right] 
\Mpl r_0^4 \right\}=0\,.
\ee
For $n=2$, there is only one branch characterized by 
\be
\phi=
\frac{2 \left[ ff'h' (3h-1)+f'^2 h (1-h)+2f'' f h (h-1) \right] 
\Mpl r_0^4 }{r^2 f^2}\,.
\label{phibra}
\ee
For $n \geq 3$, we also have the no-hair branch $\phi=0$
besides (\ref{phibra}).

Let us first consider the power $n=2$. Around the center of star, the solutions consistent with the 
boundary conditions at $r=0$ are 
\ba
f &=& f_c+\frac{f_c [\Mpl r_0^2 \rho_c (1+3w_c)
-\Mpl \phi_c^2 \alpha -24(2 w_c \rho_c r_0^2 
-\phi_c^2 \alpha)r_0^2 \phi_2 \alpha]}
{6\Mpl r_0^2 (\Mpl-16  r_0^2 \phi_2 \alpha)^2}r^2
+{\cal O}(r^3)\,,\label{fr=0a}
\\
h &=& 1-\frac{2\rho_c r_0^2+\alpha \phi_c^2}
{6 \Mpl r_0^2 (\Mpl-16 \alpha r_0^2 \phi_2)}
r^2+{\cal O}(r^3)\,,
\\
\phi &=& \phi_c+\phi_2 r^2 +{\cal O}(r^3)\,,\label{phir=0a}
\\
P &=& P_c-\frac{\rho_c(1+w_c) \{ \Mpl r_0^2 \rho_c (1+3w_c)
-\Mpl \phi_c^2 \alpha -24(2w_c \rho_c r_0^2- \phi_c^2 \alpha)
r_0^2 \phi_2 \alpha\}}
{12\Mpl r_0^2 (\Mpl-16  r_0^2 \phi_2 \alpha)^2}r^2
+{\cal O}(r^3)\,.\label{Pr=0a}
\ea
The equation to determine $\phi_2$ is given by 
\be
-2r_0^2\rho_c \left(1+3w_c\right)
+2\alpha \phi_c^2
+\frac{48r_0^2\alpha \phi_2
\left(2r_0^2w_c\rho_c-\alpha\phi_c^2\right)}
{\Mpl}
-\frac{3\left(\Mpl-16r_0^2\alpha \phi_2\right)^3\phi_c}
{2r_0^2\rho_c+\alpha\phi_c^2}
=0\,.
\label{phi2_eq}
\ee
Since Eq.~\eqref{phi2_eq} is a cubic-order equation for $\phi_2$, there are three solutions. The two of them are imaginary solutions, and the remaining one is real. 
In the limit $\alpha\to 0$, the real solution 
is approximately given by 
\begin{eqnarray}
\phi_2
&=&
\frac{1}{48\Mpl r_0^2\alpha\phi_c}
\left[
3\Mpl^2\phi_c
-2\times 6^{2/3}
\Mpl w_c \phi_c
\left(
\frac{r_0^8\rho_c^4}
        {3\Mpl^3\phi_c^2+\sqrt{3\Mpl^3\phi_c^3\left(16r_0^4w_c^3\rho_c^2+3\Mpl^3\phi_c\right)}}
\right)^{1/3}
\right.
\nonumber\\
&&
\left.
+
6^{1/3} r_0
\left\{
r_0\rho_c^2
\left(
3\Mpl^3 \phi_c^2
+
\sqrt{3
\Mpl^3\phi_c^3
\left(16r_0^4w_c^3\rho_c^2+3\Mpl^3\phi_c\right)
}
\right)
\right\}^{1/3}
\right]
+
{\cal O}
(\alpha^0).
\end{eqnarray}
Thus, the scalar field is divergent 
as $\alpha\to 0$. In general, 
in the limit $\alpha\to 0$, the same type of divergence can be also observed for $n\geq 3$. 
This indicates a pathology of the interior solutions in the power-law 
$f(R_{\rm GB}^2)$ model.

The solutions expanded far outside the star are 
\ba
f &=& 1-\frac{2M}{r}+\frac{1024M^3 r_0^6 \alpha}{r^9}
+{\cal O} (r^{-{10}})\,,\label{fr=0b} \\
h &=& 1-\frac{2M}{r}+\frac{4608 M^3 r_0^6\alpha }{r^9}
+{\cal O} (r^{-{10}})\,,
\\
\phi &=& \frac{48 M^2 \Mpl r_0^4}{r^6}
-\frac{1216512 M^4 \Mpl r_0^{10} \alpha}{r^{14}}
+{\cal O}(r^{-15})\,. \label{phir=0b}
\ea
The Schwarzschild metrics receive corrections from the coupling $\alpha$ at the order 
of $r^{-9}$. If there are NSs with a nontrivial scalar profile, the interior solutions
(\ref{fr=0a})-(\ref{phir=0a}) should be joined with the exterior solutions 
(\ref{fr=0b})-(\ref{phir=0b}).
On using the large-distance solutions (\ref{fr=0b})-(\ref{phir=0b}), 
we obtain
\be
{\cal K}=-\frac{63700992 M^6 \Mpl^6 r_0^{12} \alpha^2}
{r^{16}}+{\cal O}(r^{-17})\,.
\label{exp_k}
\ee
The leading-order term of ${\cal K}$ is negative, and hence there is a ghost instability issue at large distances.
Moreover, since ${\cal K}$ is suppressed by a high power ${\cal O}(r^{-16})$, 
it quickly approaches 0 for increasing $r$. 
Then the above solution also has a strong coupling problem  
in the asymptotic region.
Note that in Ref.~\cite{Minamitsuji:2022vbi} 
the definition of ${\cal K}$ is $2{\cal P}_1-{\cal F}$, in which case 
${\cal K} \propto r^{-18}$ as consistent with Eq.~\eqref{exp_k}.
In Ref.~\cite{Minamitsuji:2022vbi}
a small $\alpha$ expansion was used for deriving the background BH solution,
which means that our result \eqref{exp_k} is more general.

For $n \geq 3$, there is also the branch (\ref{phibra}) of 
a nonvanishing scalar field.
For this branch, the solutions in the vicinity of $r=0$ are similar 
to Eqs.~(\ref{fr=0a})-(\ref{Pr=0a}) with some modifications of 
coefficients. At large distances, the leading-order solutions are 
Schwarzschild metric components 
as in Eqs.~\eqref{fr=0b}-\eqref{phir=0b}, 
in which case we obtain
\be
{\cal K}=-\frac{3^{2n+1} \cdot 256^n M^{2(2n-1)}
\Mpl^6 n^2 (n-1)^2 r_0^{4(2n-1)}\alpha^2}{r^{4(3n-2)}}
+{\cal O}(r^{7-12n})\,.
\ee
Thus the leading-order term of ${\cal K}$ is negative, with 
a rapid decrease of ${\cal K}$ toward 0 at large distances.
To compute the quantity ${\cal K}$ above, we have not used 
the expansion with respect to a small coupling $\alpha$. 
Hence the ghost instability and asymptotic strong coupling problem of NS solutions with a nontrivial scalar profile given by the branch (\ref{phibra}) are generally present 
for an arbitrary nonvanishing coupling $\alpha$.

\section{Conclusions}
\label{consec}

In gravitational theories with a coupling to 
GB curvature invariant $R_{\rm GB}^2$, 
we studied the existence and stability of NS solutions with a nontrivial profile of the scalar field on a static and spherically symmetric background. 
For this purpose, we exploited conditions for avoiding ghost/Laplacian instabilities of 
odd- and even-parity perturbations with 
high radial and angular momentum 
modes \cite{Kase:2021mix}. 
These linear stability conditions in full 
Horndeski theories, which are summarized in Sec.~\ref{scasec}, can be applied not only to BHs, i.e., the vacuum case,
but also to NSs, i.e., the case with 
matter fluids.

The scalar-GB coupling $\alpha \xi(\phi)R_{\rm GB}^2$ gives rise to NSs endowed with a
nontrivial scalar profile both inside and outside the star. 
In Sec.~\ref{GBsec}, we derived the 
approximated background solutions
by using the expansion both around $r=0$ and 
at spatial infinity. We then studied the linear stability of them. The no-ghost condition requires that the theory has to contain 
a positive scalar kinetic term of the scalar field., i.e., $\eta>0$. 
For the existence of NS solutions with a nontrivial scalar profile free from instabilities at $r=0$, we derived the upper bound (\ref{alphacon}) on the dimensionless coupling constant $\alpha$. 
For the linear scalar-GB coupling 
$\alpha_{\rm GB} \phi R_{\rm GB}^2$, 
this bound translates to $\sqrt{|\alpha_{\rm GB}|}<0.7~{\rm km}$ to realize maximum masses 
of NSs for typical EOSs.
This theoretical limit is stronger than those observationally constrained from the GW measurements emitted from binaries 
containing NSs. 
Numerically, we confirmed that the bound (\ref{alphacon}) is sufficiently accurate for ensuring the existence of NSs with a nontrivial scalar profile consistent with all the linear stability conditions. 
As we see in Fig.~\ref{fig2}, the squared propagation speeds of odd- and even-parity perturbations deviate from 1 deep 
inside the NS, but all of them approach 1 outside the star.
Note that the propagation speeds different 
from the speed of light do not imply any acausality and instead determine the 
causal boundaries locally at each position.

In Sec.~\ref{addsec}, we incorporated several 
regular Horndeski couplings besides the scalar-GB coupling $\alpha \xi(\phi)R_{\rm GB}^2$. 
In the presence of a canonical kinetic term $\eta X$, 
the cubic-order derivative interaction of the 
scalar field $G_3 \supset \mu_3 X$ or 
the quartic-order derivative coupling 
$G_4 \supset \mu_4 X$ alone do not give rise to asymptotically-flat NS solutions with 
a nontrivial scalar profile \cite{Lehebel:2017fag}. 
We confirmed that the presence of the GB coupling is mandatory to obtain the NS solutions with a nontrivial profile of the scalar field.

In the case where the scalar-GB coupling is present, 
we showed the existence of NS solutions with a nontrivial scalar profile free from ghost/Laplacian instabilities for certain ranges of the coupling constants $\mu_3$ and $\mu_4$. 
As we see in Figs.~\ref{fig3} and \ref{fig4}, 
these new couplings do not lead to significant 
changes to the ADM mass of NSs, 
but the squared propagation speeds $c_{r3}^2$ and 
$c_{\Omega-}^2$ inside the star are subject to modifications. We also showed that NS solutions with a nontrivial scalar profile exist for 
a linear nonminimal coupling $G_4 \supset \lambda_4 \phi$ alone. Under local gravity constraints on the coupling constant $\lambda_4$, however, the scalar-GB coupling gives dominant contributions to the background scalar-field profile as well as the linear stability of NSs for $|\alpha|$ 
close to its upper limit.

In Sec.~\ref{4DEGBsec}, we addressed the linear stability of NS solutions with a nontrivial scalar profile
in regularized 4DEGB gravity. 
After the Kaluza-Klein reduction of $D$-dimensional Einstein-GB theory on a flat internal space, 
the resulting 4-dimensional action belongs to a subclass of shift-symmetric Horndeski theories.
For the solution with a nontrivial scalar 
profile (\ref{phiexp}), we showed that ${\cal K}=0$ 
and $c_{r3}^2 \to \infty$ at any radius $r$, 
and hence it is plagued by a strong coupling 
problem. Moreover the leading-order contribution 
to $c_{r3}^2/B_2$ is $-2$ at large distances, 
so there is also the Laplacian instability 
for even-parity perturbations. 
Along with the fact that the BH solutions in regularized 4DEGB gravity are also 
unstable \cite{Tsujikawa:2022lww}, 
there are no stable compact objects endowed with 
a nontrivial scalar profile in this theory.

In Sec.~\ref{fGsec}, we considered the power-law $F(R_{\rm GB}^2)$ models with the Einstein-Hilbert term, which are equivalent to 
the action (\ref{LHG2}) with $\xi(\phi)$ and $V(\phi)$ 
given by Eq.~(\ref{xiV}). 
For $n \geq 2$, there is a nonvanishing scalar-field branch 
characterized by Eq.~(\ref{phibra}).
If NSs with a nontrivial scalar profile are present, the interior solutions should join 
the large-distance exterior solutions (\ref{fr=0b})-(\ref{phir=0b}).
On using the latter, we find that the 
leading-order term of ${\cal K}$ is negative.
For increasing $r$, ${\cal K}$ rapidly decreases 
toward $-0$ with a large negative power-law 
dependence on $r$.
Hence the hairy branch (\ref{phibra}) is excluded by the problems of ghost instability and strong coupling at spatial infinity. 
This is analogous to what was found for BHs 
in the same theory \cite{Minamitsuji:2022vbi}.
In both regularized 4DEGB theory and $F(R_{\rm GB}^2)$ gravity, the instabilities of NSs with a nontrivial scalar profile arise from the unhealthy propagation 
of $\delta \phi$ associated with the absence of 
a canonical kinetic term. 
Moreover, we found that in power-law $F(R_{\rm GB}^2)$ 
models, the interior solution of the scalar field suffers from the divergence in the zero-coupling limit, 
which indicates an intrinsic pathology.

In summary, we have shown that NSs with a 
nontrivial scalar profile consistent with the 
linear stability conditions are present for the scalar-GB coupling $\alpha \xi(\phi)R_{\rm GB}^2$ besides regular Horndeski coupling functions. 
As we studied in Sec.~\ref{nonminisec}, 
nonminimal couplings with the Ricci scalar 
$G_4(\phi)R$ also give rise to NS solutions with a nontrivial scalar profile even without the scalar-GB 
coupling. The latter includes NSs with spontaneous scalarization, which can occur for a large nonminimal coupling constant of order unity \cite{Damour:1993hw,Harada:1998ge,
Novak:1998rk}. 
It will be of interest to extend our linear stability analysis to such large 
nonminimal coupling regimes by taking 
the scalar-GB coupling into account.

\section*{Acknowledgements}

MM was supported by the Portuguese national fund 
through the Funda\c{c}\~{a}o para a Ci\^encia e a Tecnologia (FCT) in the scope of the framework of the Decree-Law 57/2016 of August 29, changed by Law 57/2017 of July 19,
and the Centro de Astrof\'{\i}sica e Gravita\c c\~ao (CENTRA) through the Project~No.~UIDB/00099/2020.
MM also would like to thank Yukawa Institute for Theoretical Physics (under the Visitors Program of FY2022)
and Department of Physics of Waseda University for their hospitality.
ST was supported by the Grant-in-Aid for Scientific Research 
Fund of the JSPS Nos.~19K03854 and 22K03642.

\appendix

\section{Coefficients in the background equations}
\label{AppA}

The coefficients in Eqs.~(\ref{back1})-(\ref{back3}) are given by
\ba
&&
A_1=-h^2 (G_{3,X}-2 G_{4,\phi X} ) \phi'^2-2 G_{4,\phi} h\,,\qquad
A_2=2 h^3 ( 2 G_{4,XX}-G_{5,\phi X} ) \phi'^3-4 h^2 ( G_{4,X}-G_{5,\phi} ) \phi'\,,\notag\\
&&
A_3=-h^4G_{5,XX} \phi'^4+h^2G_{5,X}  ( 3 h-1 ) \phi'^2\,,\qquad 
A_4=h^2 ( 2 G_{4,XX}-G_{5,\phi X} ) \phi'^4+h ( 3 G_{5,\phi}-4 G_{4,X} ) \phi'^2-2 G_4\,,\notag\\
&&
A_5=-\frac12 \left[G_{5,XX} h^3{\phi'}^{5}- hG_{5,X}  ( 5 h-1 ) \phi'^3\right]\,,
\qquad
A_6=h ( G_{3,\phi}-2 G_{4,\phi\phi} ) \phi'^2+G_2\,,\notag\\
&&
A_7=-2 h^2 ( 2 G_{4,\phi X}-G_{5,\phi\phi} ) \phi'^3-4 G_{4,\phi} h\phi'\,,
\qquad
A_8=G_{5,\phi X} h^3\phi'^4-h ( 2 G_{4,X} h-G_{5,\phi} h-G_{5,\phi} ) \phi'^2-2 G_4  ( h-1 )\,,\notag\\
&&
A_9=-h ( G_{2,X}-G_{3,\phi} ) \phi'^2-G_2\,,\qquad 
A_{10}=\frac12 G_{5,\phi X} h^3\phi'^4-\frac12 h^2 ( 2 G_{4,X}-G_{5,\phi} ) \phi'^2-G_4 h\,.
\ea
The coefficients in Eq.~(\ref{Pphide}) are 
\ba
& &
\lambda_1= -\left( h'+\frac{4h}{r}+\frac{f' h}{f} \right)\phi'-2h \phi''\,,\qquad 
\lambda_2=- h \phi'^2\,,\qquad 
\lambda_3=\frac12 h \phi'^2 \left( h' \phi'+2h \phi'' \right)\,,\nonumber \\
& &
\lambda_4=\frac{2}{r^2} (1-h -r h')+\frac{hf'^2}{2f^2}
-\frac{r(2f''h+f'h')+4f' h}{2fr}\,,\nonumber \\
& &
\lambda_5=h \phi' \left[ \left( \frac{8h'}{r}+\frac{6h}{r^2}
-\frac{f'^2 h}{2f^2}+\frac{(f''r+6f')h+2rf'h'}{fr} \right)\phi'
+3h \left( \frac{f'}{f}+\frac{4}{r} \right)\phi'' \right]\,,\nonumber \\
& &
\lambda_6=h^2 \phi'^3 \left( \frac{f'}{f}+\frac{4}{r} \right)\,,\qquad 
\lambda_7=-\frac12 h^2 \phi'^3
 \left( \frac{f'}{f}+\frac{4}{r} \right) \left( h' \phi'+2h \phi'' \right)\,,
 \nonumber \\
& &
\lambda_8=\frac{1}{r^2} \left[  h'(3h-1)\phi'+2h (h - 1) \phi''  \right]
-\frac{f'^2 h^2 \phi'}{f^2 r}
+\frac{1}{fr^2} \left[ (2f'' r+3f')h^2 \phi'
+f' h(3rh'-1) \phi'+2f' h^2 r \phi'' \right]\,, \nonumber \\
& &
\lambda_9=\frac{h \phi'^2}{fr^2} \left[ f(h-1)+f' hr \right]\,,
\nonumber \\
& &
\lambda_{10}=-\frac{h \phi'^2}{2r^2} \left[ 10 h^2 \phi''
+h(7 h' \phi'-2\phi'')-h' \phi' \right]
+\frac{f'^2 h^3 \phi'^3}{2f^2 r}
-\frac{h^2 \phi'^2}{2fr^2} \left[ 
( 2f'' r+4f' )h\phi'+10f' hr \phi''+7f'h'r \phi'\right],\nonumber \\
& &
\lambda_{11}=-\frac{h^3 \phi'^4}{fr^2} (rf'+f)\,,\qquad 
\lambda_{12}=\frac{h^3 \phi'^4}{2fr^2}(rf'+f)
(h' \phi'+2h \phi'')\,.
\label{lambda}
\ea

\section{Coefficients in the perturbation equations}
\label{AppB}

The quantities appearing in the linear stability conditions (\ref{Kcon}), 
(\ref{cr3}) and (\ref{cosq}) are 
\ba
a_1&=&\sqrt{fh} \left[  \left\{ G_{4,\phi}+\frac12 h ( G_{3,X}-2 G_{4,\phi X} ) \phi'^2 \right\} r^2
+2 h \phi' \left\{ G_{4,X}-G_{5,\phi}-\frac12h ( 2 G_{4,XX}-G_{5,\phi X} ) \phi'^2 \right\} r
\right.
\notag\\
&&
\left.
+\frac12 G_{5,XX} h^3\phi'^4-\frac12 G_{5,X} h ( 3 h-1 ) \phi'^2 \right]\,, 
\nonumber \\
c_2&=&\sqrt{fh} \left[  \left\{  
\frac{1}{2f}\left( -\frac12 h ( 3 G_{3,X}-8 G_{4,\phi X} ) \phi'^2
+\frac12 h^2 ( G_{3,XX}-2 G_{4,\phi XX} ) \phi'^4
-G_{4,\phi} \right) r^2
\right.\right.
\notag\\
&&
\left.\left.
-{\frac {h\phi'}{f}} \left( 
\frac12 {h^2 ( 2 G_{4,XXX}-G_{5,\phi XX} ) \phi'^4}
-\frac12 {h ( 12 G_{4,XX}-7 G_{5,\phi X} ) \phi'^2}
+3 ( G_{4,X}-G_{5,\phi} ) \right) r
\right.\right.
\notag\\
&&
\left.\left.
+\frac{h\phi'^2}{4f}\left(
G_{5,XXX} h^3\phi'^4
- G_{5,XX} h ( 10 h-1 ) \phi'^2
+3 G_{5,X}  ( 5 h-1 ) 
\right) \right\} f'
\right.
\notag\\
&&
\left.
+\phi' \left\{ \frac12G_{2,X}-G_{3,\phi}
-\frac12 h ( G_{2,XX}-G_{3,\phi X} ) \phi'^2 \right\} r^2
\right.
\notag\\
&&
\left.
+ 2\left\{ -\frac12h ( 3 G_{3,X}-8 G_{4,\phi X} ) \phi'^2
+\frac12h^2 ( G_{3,XX}-2 G_{4,\phi XX} ) \phi'^4
-G_{4,\phi} \right\} r
\right.
\notag\\
&&
\left.
-\frac12 h^3 ( 2 G_{4,XXX}-G_{5,\phi XX} ) \phi'^5
+\frac12 h \left\{ 2\left(6 h-1\right) G_{4,XX}+\left(1-7 h\right)G_{5,\phi X} \right\} \phi'^3
- ( 3 h-1 )  ( G_{4,X}-G_{5,\phi} ) \phi' \right] \,,
\nonumber \\
c_4&=&\frac14 \frac {\sqrt {f}}{\sqrt {h}} 
\left[ {\frac {h\phi'}{f} \left\{ 
2 G_{4,X}-2 G_{5,\phi}
-h ( 2 G_{4,XX}-G_{5,\phi X} ) \phi'^2
-{\frac {h\phi'  ( 3 G_{5,X}-G_{5,XX} \phi'^2h ) }{r}} \right\}}f'
\right.
\notag\\
&&
\left.
+4 G_{4,\phi}
+2 h ( G_{3,X}-2 G_{4,\phi X} ) \phi'^2
+{\frac {4 h ( G_{4,X}-G_{5,\phi} ) \phi'-2 h^2 ( 2 G_{4,XX}-G_{5,\phi X} ) \phi'^3}{r}} \right] \,,
\ea
and
\ba
\beta_0 &=& \phi' a_1+r \sqrt{fh}{\cal H}\,,\\
\beta_1&=&\frac12 \phi'^2 \sqrt{fh} {\cal H}e_4 
-\phi' \left(\sqrt{fh}{\cal H} \right)' c_4 
+ \frac{\sqrt{fh}}{2}\left[ \left( {\frac {f'}{f}}+{\frac {h'}{h}}-\frac{2}{r} \right) {\cal H}
+{\frac {2{\cal F}}{r}} \right] \phi' c_4+{\frac {f{\cal F} {\cal G}}{2r^2}}\,,\\
\beta_2&=& \left[ \frac{\sqrt{fh}{\cal F}}{r^2} \left( 2 hr\phi'^2c_4
+\frac{r \phi' f' \sqrt{h}}{2\sqrt{f}}{\cal H}-\phi' \sqrt{fh}{\cal G} \right)
-\frac{\phi' fh {\cal G}{\cal H}}{r} \left( \frac{{\cal G}'}{{\cal G}}
-\frac{{\cal H}'}{{\cal H}}+\frac{f'}{2f}-\frac{1}{r} \right) \right]a_1 
-\frac{2}{r} (fh)^{3/2}{\cal F}{\cal G}{\cal H}\,, \qquad\,\,\\
\beta_3&=& \frac{\sqrt{fh}{\cal H}}{2}\phi'  
\left( hc_4'+\frac12 h' c_4-\frac{d_3}{2} \right) 
-\frac{\sqrt{fh}}{2} \left( \frac{\cal H}{r}+{\cal H}' \right) 
\left( 2 h \phi'c_4+\frac{\sqrt{fh}{\cal G}}{2r}
+\frac{f'\sqrt{h}{\cal H}}{4\sqrt{f}} \right)\notag\\
&&
+{\frac {\sqrt {fh}{\cal F}}{4r} \left(  2 h \phi'c_4
+\frac{3\sqrt{fh}{\cal G}}{r}
+\frac{f'\sqrt{h}{\cal H}}{2\sqrt{f}}
 \right) }\,,
 \ea
with 
\ba
e_4 &=&{\frac {1}{\phi'}}c_4'-{\frac {f'}{4f h \phi'^2}} 
\left( \sqrt{fh} {\cal H} \right)'
-{\frac {\sqrt {f}}{2\phi'^2\sqrt {h}r}}{\cal G}'
+{\frac {1}{h\phi' r^2} \left( {\frac {\phi''}{\phi'}}+\frac12 {\frac {h'}{h}} \right) }a_1
\notag\\
&&
+{\frac {\sqrt{f}}{8\sqrt{h}\phi'^2} 
\left[ {\frac { ( f' r-6 f ) f'}{f^2r}}
+\frac {h' ( f' r+4 f ) }{fhr}
-{\frac {4f ( 2 \phi'' h+h' \phi')}
{\phi' h^2r ( f' r-2 f ) }} \right] }{\cal H}
+{\frac {h'}{2h\phi'}}c_4
-\frac{f'r-2f}{4\sqrt{fh}r\phi'}
\frac{\partial {\cal H}}
{\partial \phi}
\notag\\
&&
+{\frac {f' hr-f}{2r^2\sqrt {f}{h}^{3/2}\phi'^2}}{\cal F}
+{\frac {\sqrt {f}}{2r\phi'^2{h}^{3/2}} 
\left[ {\frac {f ( 2 \phi'' h+h' \phi' ) }{h\phi'  ( f' r-2 f ) }}
+{\frac {2 f-f' hr}{2fr}} \right] }{\cal G}
+\frac{\sqrt{f} (\rho+P)}{2h^{3/2} \phi'^2}
\,, \label{e4}\\
d_3&=&
-{\frac {1}{r^2} \left( {\frac {2\phi''}{\phi'}}+{\frac {h'}{h}} \right) }a_1
+{\frac {f^{3/2}h^{1/2}}{ ( f' r-2 f ) \phi'} \left( 
{\frac {2\phi''}{h\phi' r}}
+ {\frac {{f'}^{2}}{f^2}}
- {\frac {f' h'}{fh}}
-{\frac {2f'}{fr}}
+{\frac {2h'}{hr}}
+ {\frac {h'}{h^2r}} \right) }{\cal H}
\notag\\
&&
+\frac{f'r-2f}{2r} \sqrt{\frac{h}{f}} 
\frac{\partial{\cal H}}{\partial \phi}
+{\frac {\sqrt {f}}{\phi' \sqrt {h}r^2}}{\cal F}
-{\frac {{f}^{3/2}}{\sqrt {h} ( f' r-2 f ) \phi'} 
\left( {\frac {f'}{fr}}+{\frac {2\phi''}{\phi' r}}+{\frac {h'}{hr}}-\frac{2}{r^2} \right) }{\cal G}-\frac{\sqrt{f} (\rho+P)}{\phi' \sqrt{h}} \,.
\ea

\bibliographystyle{mybibstyle}
\bibliography{bib}

\end{document}